\documentclass[preprint,superscriptaddress,
amssymb,
aps,floatfix]{revtex4-2}

\usepackage{bm,amsmath,color,bm}
\usepackage{graphicx,subfigure}
\usepackage{dcolumn}
\usepackage{epsfig}
\usepackage{latexsym}
\usepackage{amsfonts}
\usepackage{amssymb}
\usepackage{mathrsfs}
\usepackage{lineno}
\usepackage{siunitx}
\usepackage{hyperref}
\usepackage{soul}

\usepackage{comment}

\begin{document}

\title{Elucidating chirality transfer in liquid crystals of viruses
}

\author{Eric Grelet}
\email[]{eric.grelet@crpp.cnrs.fr}
\affiliation{Univ. Bordeaux, CNRS, CRPP, UMR 5031, F-33600 Pessac, France\\}

\author{Maxime M. C. Tortora}
\email[]{tortora@usc.edu}
\affiliation{Univ. Claude Bernard Lyon 1, ENS de Lyon, CNRS, LBMC, UMR 5239, Inserm U 1293, 
F-69007 Lyon, France}
\affiliation{Current address: University of Southern California, Department of Quantitative and Computational Biology, 
Los Angeles, CA 90089, USA}

\date{\today}

\begin{abstract}

Chirality is ubiquitous in nature across all length scales, with major implications spanning the fields of biology, chemistry and physics to materials science. 
How chirality propagates from nanoscale building blocks to meso- and macroscopic helical structures remains an open issue. Here, working with a canonical system of filamentous viruses, we demonstrate that their self-assembly into chiral liquid crystal phases quantitatively results from the interplay between two main mechanisms of chirality transfer: electrostatic interactions from the helical charge patterns on the virus surface, and fluctuation-based helical deformations leading to viral backbone helicity. Our experimental and theoretical approach provides a comprehensive framework for deciphering how chirality is hierarchically and quantitatively propagated across spatial scales. 
Our work highlights the ways in which supramolecular helicity may arise from subtle chiral contributions of opposite handedness which either act cooperatively or competitively, thus accounting for the multiplicity of chiral behaviors observed for nearly identical molecular systems. 

\end{abstract}

\maketitle 
\clearpage

\section{Introduction}


Understanding and controlling the propagation 
of chirality across length scales, 
from chiral molecular primary units 
such as molecules possessing asymmetric carbons 
to ordered helical superstructures and chiral bulk assemblies, 
is of paramount importance 
in multiple 
contexts 
encompassing the fields of biology, chemistry and physics to nanotechnology and materials science \cite{Liu2015,
Morrow2017,Nemati2018,Nemati2022,Zhang2022,Sang2022,Kotov2022}. 
Among large-scale chiral patterns, the liquid-crystalline organization known as the \textit{cholesteric phase}
can be arguably considered as the quintessential helical assembly (Extended Data Fig.~\ref{Handedness}). Beyond widespread technological applications ranging from the display industry 
to smart windows \cite{Mitov2012,Geng2022,
Bisoyi2022}, cholesteric structures are also ubiquitously found in 
biological matter --- both \textit{in vivo}, as found in some plant tissues,
the cuticles of arthropods such as beetles and crabs \cite{Mitov2017}, and also \textit{in vitro} in solutions of cholesterol derivatives \cite{Reinitzer1888}, 
nucleic acids \cite{Livolant1996,Zanchetta2010,Siavashpouri2017}, viruses \cite{Dogic2000,Grelet2003,Tombolato2006}, amyloid fibrils \cite{Bagnani2019}, chitin \cite{Belamie2004}, cellulose nanocrystals \cite{Araki2001b,Usov2015,Honorato-Rios2020,Parton2022}, etc.

Despite considerable efforts 
over past decades \cite{Straley1976,Harris1999,Osipov1988,Grelet2003,Tombolato2006,Cherstvy2008,Dussi2016,Siavashpouri2017,Tortora2020,Parton2022}, the mechanisms responsible for the hierarchical 
propagation of chirality --- i.e., the causal relationship between the microscopic properties of molecular building blocks and their emergent macroscopic structure 
 --- remain largely unresolved. 
The underlying difficulty 
is 
that chiral interactions are 
intrinsically weak. This can be illustrated by the fact that the preferred mutual twist angle 
between two adjacent particles found in standard cholesteric arrangements is typically a fraction of a degree. This should be compared to the average angle by which the rods locally fluctuate in such liquid-crystalline phases, 
which is typically of the order of tens of degrees, two orders of magnitude larger \cite{Dogic&Fraden}. Consequently, any model of the system that aims at predicting the hierarchical transfer of chirality from the molecular level to larger helical structures requires a highly accurate description of the chiral interactions in order to reliably account for the magnitude and sense of the resulting helical periodicity. 

In this work, we study the cholesteric liquid crystalline phase formed by aqueous suspensions of filamentous viruses both experimentally and theoretically. These viruses, known as bacteriophages for their ability to infect bacteria, are widely used 
as a model system in genetic engineering due to their ease of modification \cite{Smith1997,Marvin2006,Lee2009,Marvin2014}, 
in soft condensed matter as monodisperse rod-like model particles \cite{Dogic&Fraden,Gibaud2012,Grelet2014}, and in nanotechnology as versatile and functionalizable colloidal templates \cite{Willis2008,Lee2009,Chung2011}. Here, we demonstrate that their self-organization 
into the cholesteric liquid crystalline state 
quantitatively results from the interplay between three different contributions of chirality transfer: 
1) steric repulsion between the screw-shaped capsids arising from the helical
arrangement of coat proteins on the virus surface; 
2) electrostatic interactions between the helical charge pattern on the virus surface;
3) long-wavelength chiral deformations stemming from the virus flexibility and leading to a 
coherent supramolecular helical morphology of the virus backbone (Fig.~\ref{SchemePymol}).

These sources of chirality transfer are independently probed by tuning the ionic environment and by the chemical functionalization 
of two specific virus strains that form chiral nematic phases with opposite handedness. They are confirmed by numerical models including a detailed atomistic description of the inter-particle force field accounting quantitatively for the full set of our experimental results. 

\section{Results and discussion}

\begin{figure}
	\includegraphics[width=0.75\columnwidth]{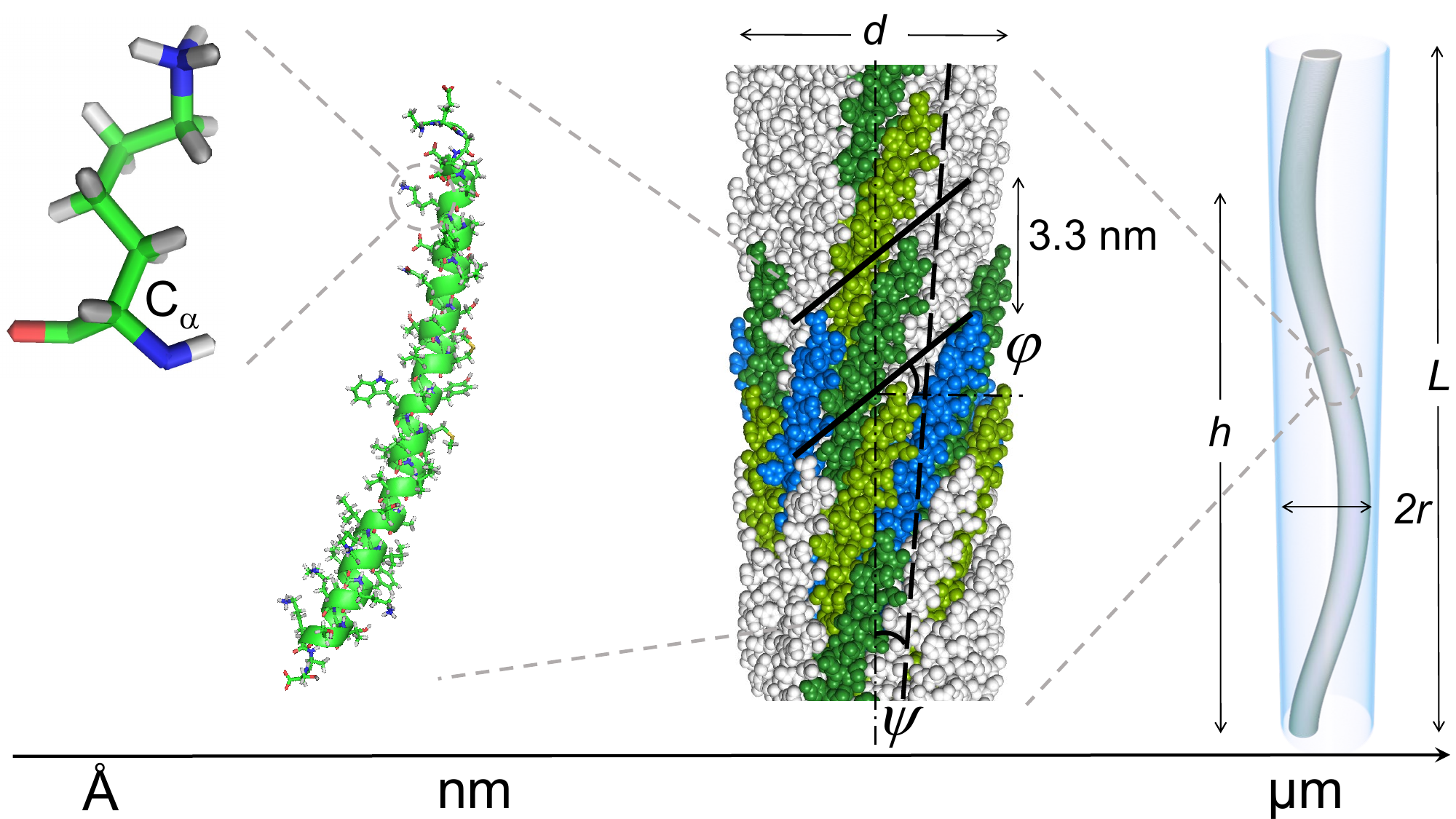}
	\caption{\textbf{Structural chirality of filamentous viruses at 
 various hierarchical scales}, ranging from the asymmetry of
		C$_\alpha$ atoms of the amino acids 
  of the main coat proteins p8, to the $\alpha$-helical
		structure of these proteins and their 
        helical arrangement on the virion surface. At the micrometer scale, the flexibility of the virus 
        suggests the existence of 
        suprahelical backbone deformation modes of radius $r$ and internal helical pitch $h$, stemming from long-wavelength chiral
		fluctuations of the whole virus shape. The atomistic model of both M13 and Y21M filamentous viruses are respectively derived from the 
		 1IFI and 2C0W capsid structures 
      \cite{Marvin2006,Marvin2014},
		 and deposited in the Protein Data Bank (PDB). A primary right-handed thread angle 
   resulting from the main groove formed by the major coat proteins p8 
		and indicated by black lines is found to be $\varphi < 45^{{\rm o}}$ for both mutants. 
        A secondary groove and 
        associated thread can be identified  (black long-dashed lines) which has a right-handed helical symmetry for the M13 capsid ($\psi = 5.15^{{\rm o}}$, 1IFI model, as shown in the representation here) but is achiral for Y21M 
        ($\psi = 0^{{\rm o}}$, 2C0W model). Different colors are used to highlight a few main coat proteins p8 and the associated symmetries  of the capsid. 
        } 
	 
	\label{SchemePymol}
\end{figure}

\begin{figure}
	\includegraphics[width=0.7\columnwidth]{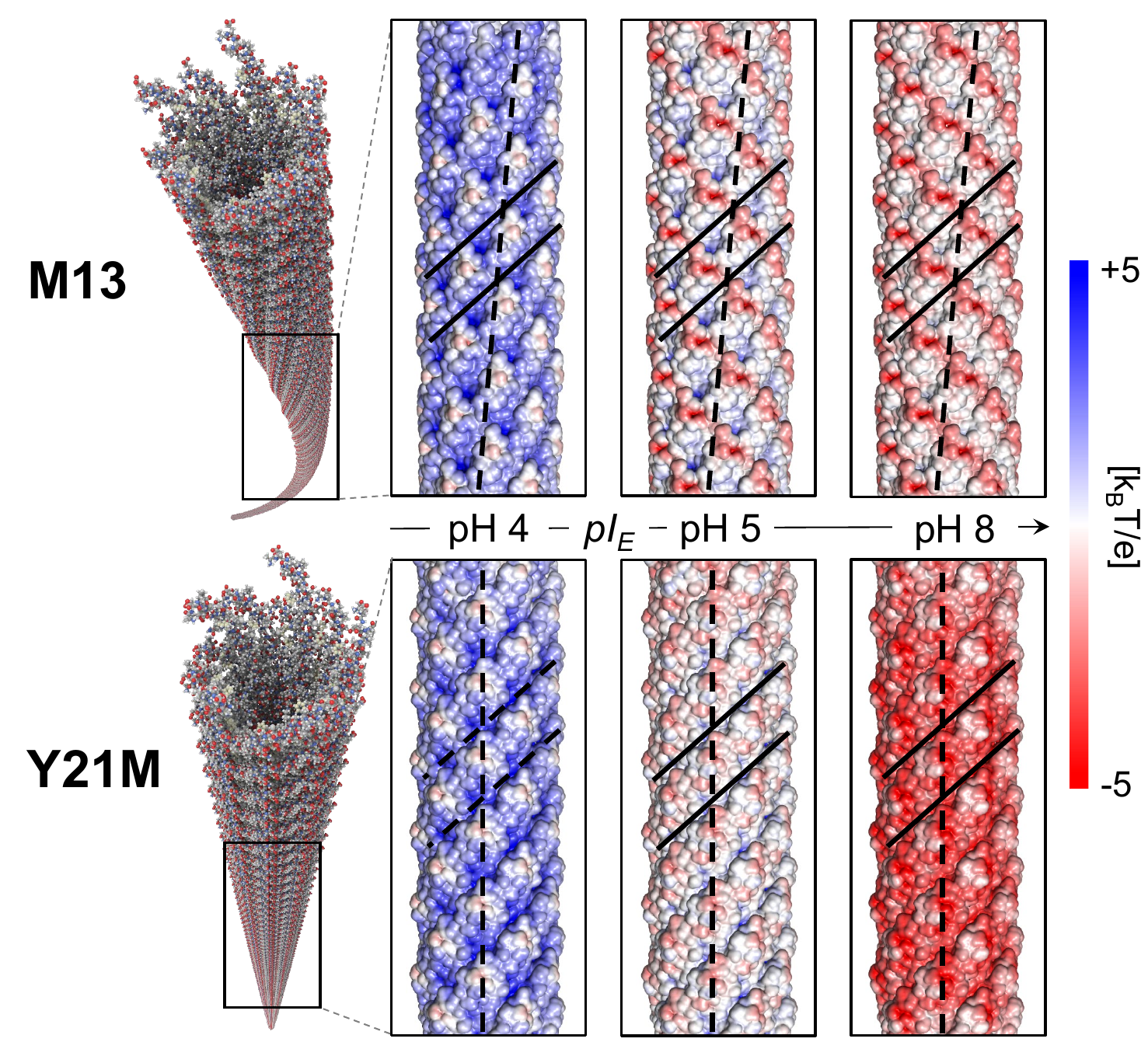}
	\caption{\textbf{
        Atomistic representation of the semi-flexible M13 and stiff Y21M virions and pH dependence of their charge distribution} calculated using Advanced Poisson-Boltzmann Solver (APBS). 
		The electrostatic surface potential of both viruses, rendered using the PyMOL software's APBS plugin~\cite{Jurrus2018}, 
		is shown to vary from negative (red) to positive (blue) by decreasing the pH below the virus isoelectric point $pI_E$. 
		The two main threads identified on the virus capsid (as defined in Fig. \ref{SchemePymol}) are indicated by continuous and long-dashed black lines, respectively. 
}
	\label{PyMol-Helix&ChargeDistrib}
\end{figure}

Filamentous bacteriophages of the Ff family 
are 
micrometer-long semi-flexible particles, 
which are primarily composed of a single-stranded DNA around which about 3000 copies of the main coat protein p8 are 
assembled in an overlapping, interdigitated helical
structure (Fig. \ref{SchemePymol}). 
Two closely-related strains are used, M13 and Y21M, which are structurally and biologically very similar \cite{Marvin2006} and which bear a net negative charge in physiological conditions (Fig.~\ref{PyMol-Helix&ChargeDistrib}).
The primary difference, a single amino acid mutation occurring in 
the central region of the 50 amino-acid long coat protein p8, i.e. at position 21, slightly alters the symmetry of the phage capsid (see Methods), as shown by Marvin \textit{et al.} with high-resolution X-ray diffraction \cite{Marvin2006,Marvin2014}. The resulting 3D
atomistic structure of the M13 and Y21M virus capsids are deposited in the Protein Data Bank (PDB) under 1IFI and 2C0W models, respectively. 
\par
\begin{figure}
	\includegraphics[width=0.7\columnwidth]{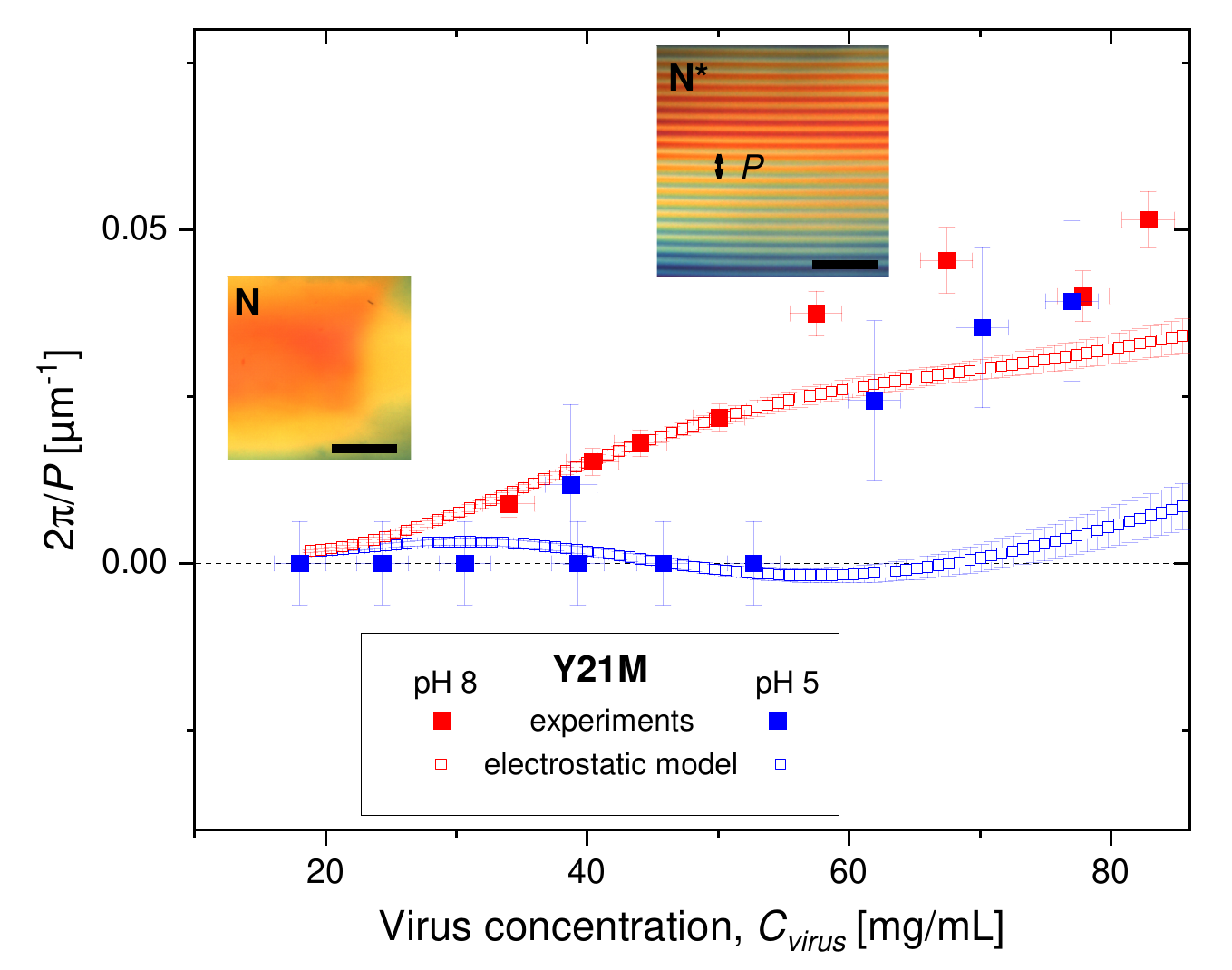}
	\caption{\textbf{Cholesteric to nematic 
        crossover 
        by decreasing charge fully accounted by the electrostatic model in Y21M virus suspensions.}  
		At low virus charge, i.e., at pH~5 
		close to $pI_E$ (blue full symbols), the cholesteric pitch unwinds by decreasing virus concentration until reaching the nematic phase (2$\pi/P$=0), as illustrated by the two polarization micrographs showing, respectively, the absence in the nematic (left inset) and the  presence in the cholesteric (right inset) of fingerprints (Scale bars: 250~$\mu m$). By contrast, for highly charged viral rods at pH~8 (red full symbols), 
		a cholesteric phase of smaller pitch $\vert P \vert$ is observed across the whole liquid crystalline range. 
        For both conditions of pH, the ionic strength is fixed at $I_S$=110~mM. The electrostatic model (open symbols) 
        \textit{quantitatively} accounts for the experimental results of stiff Y21M in the dilute range of the cholesteric organization both at low and high charges (pH~5 and 8, respectively).
	}
	\label{Y21M&M13vsElectroModel}
\end{figure}

Both capsids display a main right-handed groove (Fig.~\ref{SchemePymol}) stemming from the helicoidal wrapping of the major coat proteins p8, whose thread angles differ, $\varphi = 39.85^{{\rm o}}$ for M13 and $43.15^{{\rm o}}$ for Y21M. A secondary thread can be identified  which has a right-handed helical symmetry for the M13 capsid ($\psi = 5.15^{{\rm o}}$, 1IFI model) but is achiral for Y21M due to its exact two-fold screw symmetry ($\psi = 0^{{\rm o}}$, 2C0W model). These subtle structural differences between the two viruses result in a large 
change in 
their stiffness (see Methods): the M13 strain is semi-flexible with a persistence length $L_p$ over contour length $L_c$ ratio of $L_p/L_c~\simeq 3$ whereas Y21M is nearly rigid with $L_p/L_c \simeq 11$ (Fig.~\ref{PyMol-Helix&ChargeDistrib}) \cite{Barry2009}. 
When dispersed in buffer solutions of controlled pH and ionic strength $I_S$ (see Methods), both rod-shaped virions exhibit the same liquid crystal phase sequence \cite{Dogic2000,Grelet2014}, and self-organize into the cholesteric phase where the helical twist occurs perpendicular to the average particle orientation 
as shown in Extended Data Fig.~\ref{Handedness}. The cholesteric ordering is characterized by its helical periodicity, or pitch $P$, which is defined as positive (negative) for right (left) handedness. It is worth mentioning that, despite the similarity in the structure of the phages, their cholesteric phases have opposite handedness, left-handed for M13 and right-handed for Y21M (Extended Data Fig.~\ref{Handedness}) \cite{Tombolato2006,Barry2009}. By varying the ionic conditions, the cholesteric pitch $\vert P \vert$ of both viral mutants increases with increasing the ionic strength $I_S$ \cite{Dogic2000,Tombolato2006} (Extended Data Fig.~\ref{Y12M&M13rawData_pH8}) and as pH decreases towards the isoelectric point of the virus, $pI_E$ 
(Figs.~\ref{PyMol-Helix&ChargeDistrib}--\ref{Y21M&M13vsElectroModel}). This sensitivity of the cholesteric pitch to modulations of either the range or the intensity of electrostatic interactions
shows their major contributions 
in the cholesteric assembly. 

Different mechanisms have been suggested to predict the value and sense of the cholesteric pitch from the molecular features of the chiral 
constituents. 
They mostly rely on two main classes of chiral intermolecular potentials: 
steric interactions based on hard-core repulsion at different length scales, ranging from local helical threads decorating rod-like particles to full 
helical
shapes such as hard helices \cite{Straley1976,Frezza2014,Dussi2015,Dussi2016}, and 
electrostatic interactions resulting from the helical charge distributions carried by the particles \cite{Kornyshev2002,Cherstvy2008,Tombolato2006}, which have been further generalized to also include chiral dispersion forces \cite{Osipov1988,Wensink2009}. However, most of these theoretical approaches have been developed at a 
coarse-grained level aiming both at capturing the key-physical mechanisms and 
keeping the calculations tractable. As a consequence, their limited accuracy 
may be insufficient to fully account for the complexity of real experimental systems, 
and does generally not provide a quantitative agreement between model predictions and experimental measurements \cite{Harris1999,Tombolato2006}.

Following the seminal work of Onsager \cite{Onsager1949}, Straley was the first to propose a microscopic
theory of the cholesteric phase based on excluded volume interaction by considering rod-like particles exhibiting additional chiral threads similar to
screws \cite{Straley1976}. The steric hindrance between two
screw-like rods is minimal not when they are parallel to each other, but when they
approach each other at a specific angle at which the chiral grooves can interpenetrate (Extended Data Fig.~\ref{screws}). 
Depending on the thread angle $\varphi$, the helical twist resulting from the optimal packing of right-handed screws may be either right-handed if $\varphi<45^{{\rm o}}$, 
or left-handed if $\varphi>45^{{\rm o}}$. 
This simple example illustrates the non-trivial relationship between the handedness of microscopic building blocks and that of their macroscopic helical assemblies, which can thus yield either homo- ($\varphi<45^{{\rm o}}$) or hetero-chiral ($\varphi>45^{{\rm o}}$) structures depending on their detailed molecular morphology (Extended Data Fig.~\ref{screws}).
Such steric-based chiral interactions have been already observed in experimental systems, for instance between helical nanofilaments of the B4 liquid crystalline phase \cite{Zhang2014}.  
As most biopolymers display helical charge patterns, models including electrostatic interactions have also been developed, but have usually 
been limited to 
coarse-grained descriptions
~\cite{Kornyshev2002,Tombolato2006,Wensink2009}.  

Here, we go beyond coarse-grained approaches and 
we introduce an \textit{electrostatic model} built at the atomistic level and based on an all-atom representation of the ground-state conformation of the virus, in the absence of thermal fluctuations (Fig.~\ref{SchemePymol}).  
Our model provides an atomic-level description of the virion pair interaction energy $U_{\rm inter}$ which explicitly accounts for the screened electrostatic, steric and van der Waals forces involving each of the $\sim 3\times 10^6$~atoms 
within the virus capsid (Fig.~\ref{PyMol-Helix&ChargeDistrib}). Potential parameters are set per amino acid and atom type using an iteration of the GROMOS force field optimized to reproduce the free enthalpy of solvation of biomolecular compounds, which includes a dependence on ionic content in the form of implicit-solvent electrostatic interactions (see Methods) \cite{oostenbrink2004}. 
The atomistic representations of the Y21M and M13 capsids are respectively reconstructed from their 2C0W and 1IFI three-dimensional structures. 
The corresponding protonation states and charge distributions are inferred at various pH conditions using 
standard protein structure preparation software (Fig.~\ref{PyMol-Helix&ChargeDistrib}, see Methods). As thermal fluctuations of the virion backbone are neglected, the electrostatic model is effectively developed in the high stiffness limit of the filamentous virus ($L_p \longrightarrow \infty$).
The cholesteric pitch $P$ and twist elastic modulus
$K_{22}$ are then obtained from the minimization of the associated free energy  $\mathscr{F}$ (see Methods and Supplementary Sections~I~\&~II), and the results are shown in Fig.~\ref{Y21M&M13vsElectroModel}. An excellent quantitative agreement is found between theory and experiments for Y21M virion, at both low and high pH. The electrostatic model captures not only the sense and magnitude of the Y21M cholesteric helicity, but also its unwinding when the virus surface charge decreases. For the highest virus concentrations, the model slightly underestimates the resulting helicity of the system (Fig.~\ref{Y21M&M13vsElectroModel}). This is likely arising from the second-virial approximation underlying our model
, which is expected to be increasingly inaccurate at higher virus volume fractions (see Methods). 
\par
However, the electrostatic model does not account for the chiral phase behavior of the M13 phage (see Supplementary Section~III),
for which it 
strongly underestimates
the chirality of the system. 
Thus, although the electrostatic model succeeds in capturing the experimentally-observed cholesteric assembly of Y21M virions, another mechanism of chirality transfer has to be invoked in the case of M13 suspensions. 

\begin{figure}
	\includegraphics[width=
 0.65\columnwidth]{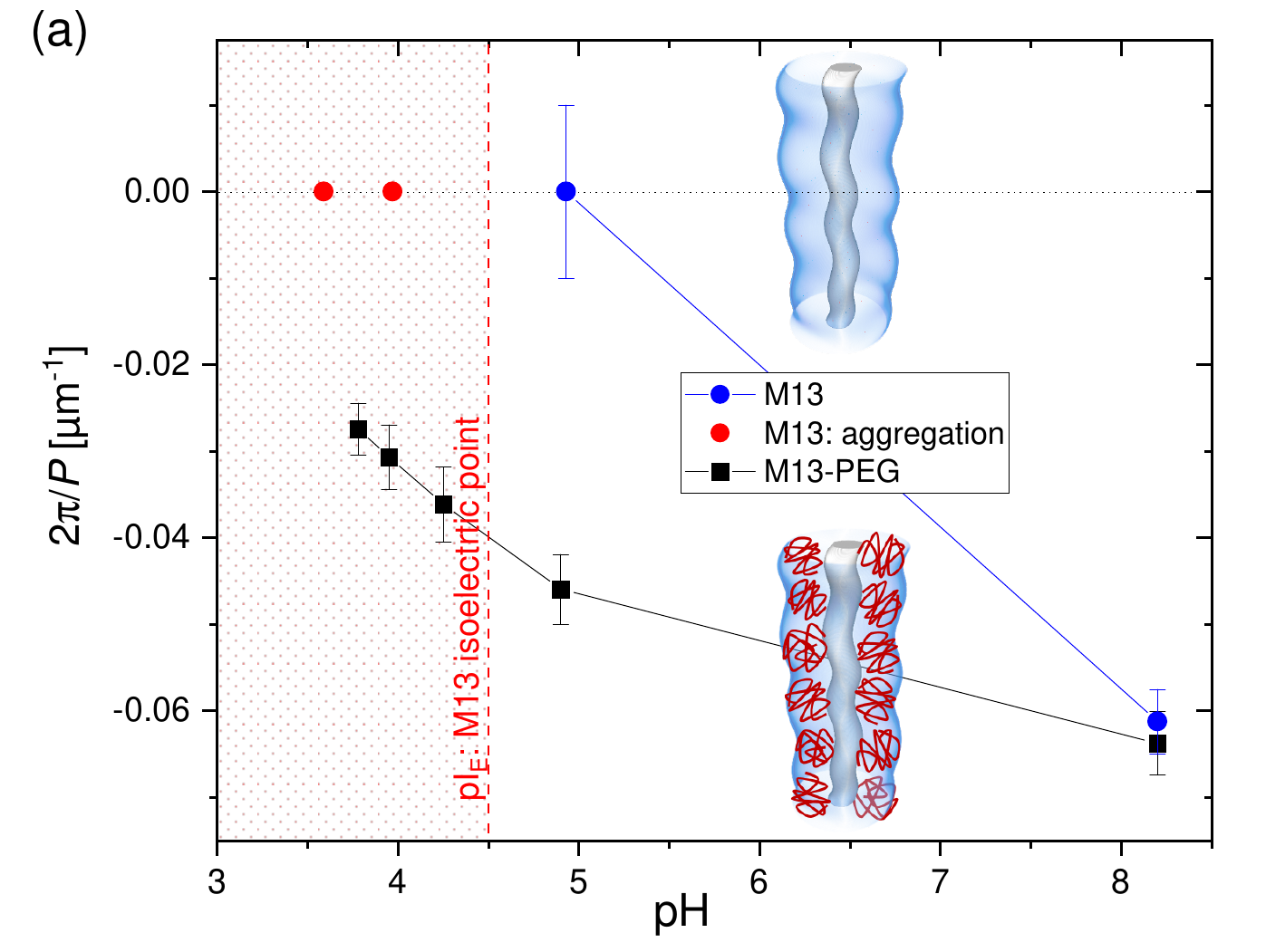}
	\includegraphics[width=
 0.64\columnwidth]{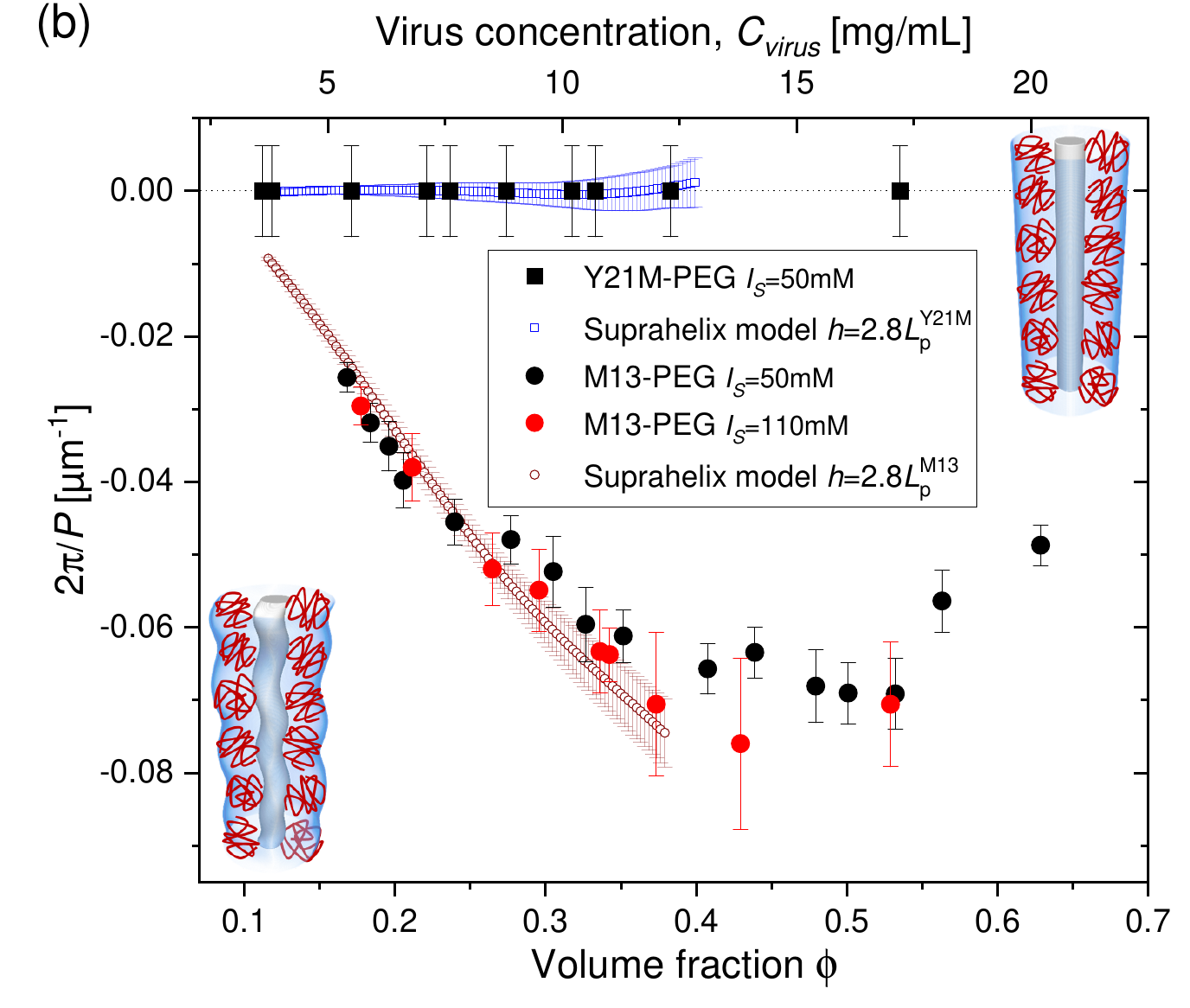}
	\caption{\textbf{Cholesteric ordering in sterically stabilized semi-flexible viruses and suprahelix model. 
    (a) pH dependence of the inverse of the cholesteric pitch \textit{P}} in M13 
    ($C_{M13}=42$~mg/mL) and 
    M13-PEG 
    ($C_{M13-PEG}=10.5$~mg/mL) 
		 suspensions at $I_S$=110~mM. Each concentration is chosen in the fully liquid crystalline regime such that $C_{virus}/C_{iso}\simeq 2$, where $C_{iso}$ is the isotropic binodal concentration. 
		As the pH is decreased below the isoelectric point $pI_E$ 
    (red dotted area), colloidal aggregation is observed in pristine M13 suspensions (red symbols). In contrast, sterically stabilized semi-flexible M13-PEG virions (full black squares) still exhibit a left-handed cholesteric phase at, and below, the isoelectric point $pI_E$. 
  }
\end{figure}
\addtocounter{figure}{-1}
\begin{figure}
 \caption{\textbf{(b) Influence of viral rod flexibility on the inverse cholesteric pitch \textit{P} in PEGylated virus suspensions.}
	The semi-flexible M13-PEG system with a persistence length $L_p^{M13} \simeq 3 L_c$  (with $L_c$ the virion contour length, as $ L_c \simeq L_c^{M13}\simeq L_c^{Y21M}$) exhibits a cholesteric phase whose pitch (full circles) is mostly independent of ionic strength $I_S$ at high enough salt concentration, whereas the stiff Y21M-PEG ($L_p^{Y21M} \simeq11 L_c$) has a nematic-like behavior (square symbols) without any observable chirality propagation. Our model of 
 suprahelical backbone deformations with an internal pitch of $h=2.8L_p$, 
    as described in the main text, is able to account \textit{quantitatively} for the chiral behavior of stiff 
	and semi-flexible 
	PEGylated viruses. }
	\label{SupraHelix}
\end{figure}

To further investigate the nature of the chiral interactions between viruses,
a shell of neutral hydrophilic polymers (PEG, see Methods) is covalently grafted on both virion surfaces (Extended Data Fig.~\ref{Texture}). The phase behavior of the polymer coated viruses (M13-PEG and Y21M-PEG) becomes independent of ionic strength at high salt concentration \cite{Grelet2016}. In contrast to pristine viruses, whose colloidal stability relies on electrostatic repulsion and therefore aggregate at the isoelectric point $pI_E$, PEGylated viruses are sterically stabilized and can then be studied at varying pH (Extended Data Fig.~\ref{Texture}). This includes the range close to the isoelectric point $pI_E$, where no net electric charge remains on the virus surface 
and therefore the electrostatic interactions vanish. Interestingly, in these conditions, a left-handed cholesteric phase still persists for M13-PEG system at $pI_E$, whereas Y21M-PEG suspensions exhibit a nematic phase as expected in absence of electrostatic interactions (Fig.~\ref{SupraHelix}). This implies that another general mechanism driving the chirality transfer for the M13 and M13-PEG phages needs to be considered. This independence from the electrostatic interactions leads to a cholesteric pitch which does not depend on ionic strength, as shown in Fig.~\ref{SupraHelix}b. 
Furthermore, the cholesteric pitch found for M13 and M13-PEG are nearly identical when the colloidal stability is preserved (i.e. at pH~8, see Fig.~\ref{SupraHelix}a) and the virus concentration is rescaled by the binodal concentration associated with the stability limit of the isotropic phase
$C_{iso}$. This indicates that the mechanism of chirality transfer for M13 and M13-PEG to the cholesteric ordering is not sensitive to the structural details and symmetries of the phage surface, and occurs at a much higher length scale than the atomic level. A key difference between the two virus strains is their stiffness, M13 is semi-flexible whereas Y21M is a stiff colloidal rod (see Methods). Together, these observations suggest a mechanism of chirality transfer based on long-wavelength helical fluctuations of the virus backbone, which we will henceforth 
refer to as the \textit{suprahelix model} \cite{Grelet2003,Tortora2020}.

In detail, the suprahelix model accounts for the 
ability of the virion flexibility to promote long-wavelength chiral deformation modes due to thermal fluctuations, leading to a coherent helical morphology spanning the entire virus shape \cite{Grelet2003}. While no 
primary proof of such helical conformation has been reported yet \cite{Willis2008}, we provide here indirect experimental evidence for the formation of helical supramolecular self-assemblies from cholesteric suspensions of filamentous viruses under depletion interactions (Extended Data Fig.~\ref{HelicalBundles}). These 
helical structures are reminiscent of 
the helical morphology  
assumed to arise from the backbone fluctuations of individual viruses in the suprahelix model, and therefore strongly support this hypothesis. 
As the full mechanical modeling of filamentous phages over experimentally-relevant timescales remains computationally inaccessible to standard atomistic simulations, we approximate these helical deformation modes by a \textit{mean} effective backbone conformation, described by a hard helix of radius $r$ and internal pitch $h$ (Fig.~\ref{SchemePymol}). In this framework, $h>>L_c$ indicates that the internal helical periodicity may far exceed the full length of an individual virus, giving rise to a weakly-curved, ``suprahelical" higher-order shape \cite{Frezza2014}.  
\par
Although the link between ground-state and fluctuation-induced chirality is generally non-trivial~\cite{Tortora2020}, the left-handed phases displayed by the more flexible M13 phages (Extended Data Fig.~\ref{Handedness}) would suggest that the corresponding backbone deformations should be predominantly right-handed --- in agreement with simple geometric arguments governing the self-assembly of weakly-curled helices (Fig.~\ref{SchemePymol}, Extended Data Fig.~\ref{screws}). Based on the ``tube'' model of polymer deflection 
~\cite{Odijk1986}, it is shown in Supplementary Section~IV that the resulting suprahelical conformation may then be expressed in terms of the internal pitch $h$ as the \textit{sole} adjustable parameter for a given virion persistence length $L_p$. 
The cholesteric pitch $P$ and the twist elastic constants $K_{22}$ are also obtained for this model from the minimization of the free energy $\mathscr{F}$ associated with these weakly-curled suprahelices interacting pairwise through hard-core repulsion. 

\begin{figure}
	\includegraphics[width=0.7\columnwidth]{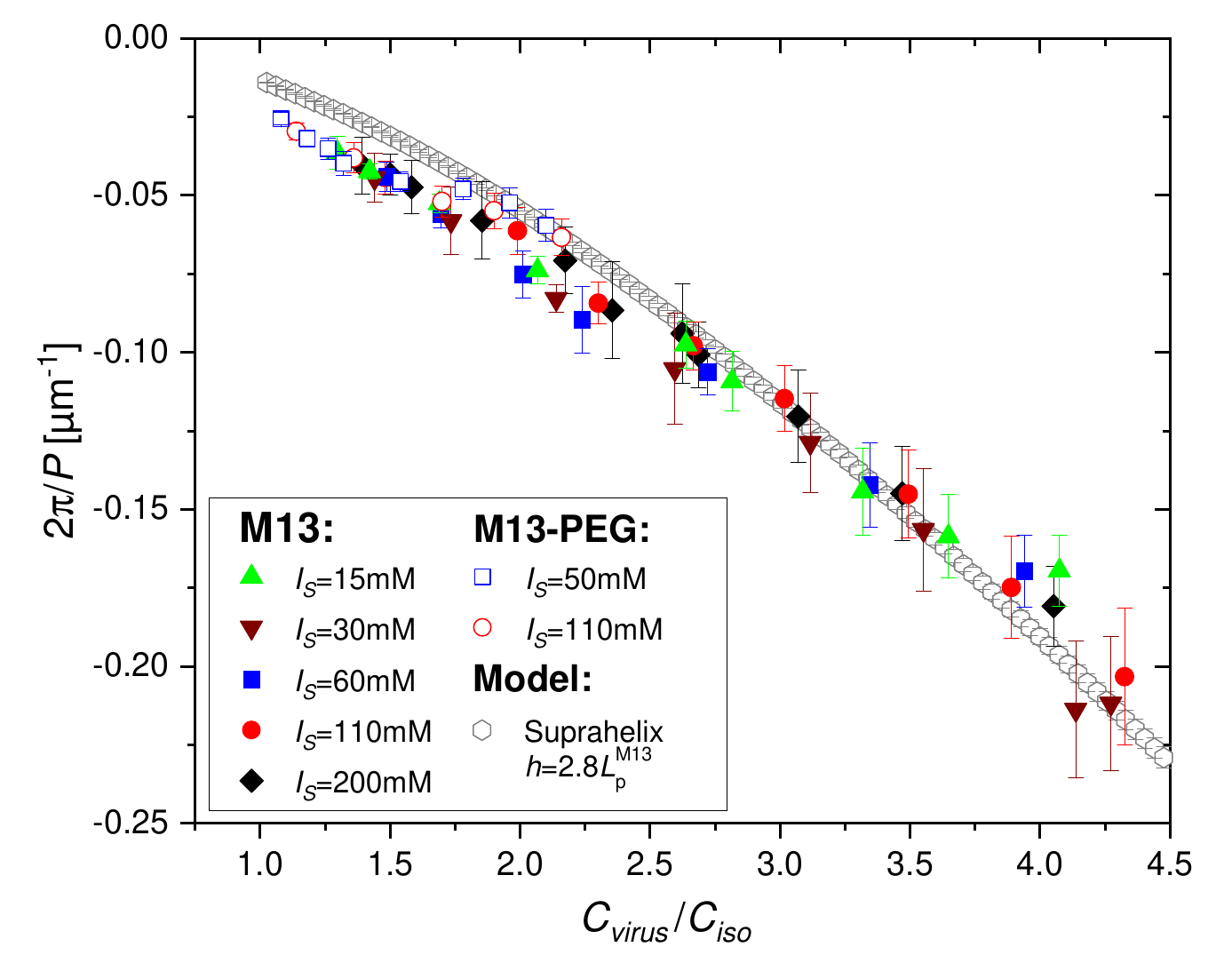}
	\caption{\textbf{Master curve of the cholesteric pitch \textit{P} for semi-flexible viruses accounted by the suprahelix model.} 
 After renormalization of the particle concentration by the binodal value corresponding to the stability limit of the isotropic phase
 $C_{iso}$, the inverse cholesteric pitch $P$ of both PEGylated M13-PEG limited to the dilute regime, 
 and of charged M13 for all probed ionic strengths (see raw data in Extended Data Fig.~\ref{Y12M&M13rawData_pH8}), 
 collapse onto a unique master curve. All renormalized data are accounted for by the suprahelix model with an internal helical pitch of $h=2.8L_p$. 
 The binodal value C$_{iso}$ used for the normalization of the theoretical model is corrected for rod flexibility \cite{Chen1993}.
 Since $C_{iso}$ is solely a function of the particle features $(d,L_c,L_p)$~\cite{Onsager1949,Chen1993}, this indicates 
that the mechanism of chirality transfer of semi-flexible virions is mostly entropy driven, as explicitly demonstrated in Supplementary Section~V.  
	}
	\label{MasterCurve}
\end{figure}

The results are shown in Figs.~\ref{SupraHelix}b--\ref{MasterCurve}, where the internal pitch $h$ of the suprahelical conformation has been set such that
$h=2.8L_p$. The PEGylation of viruses is limited to about 10\% of the capsid proteins, so that 
the grafted polymers may only change the surface properties of the viruses and do not alter their internal structure (see Methods), 
leading to 
the same persistence length --- and therefore similar helical 
deformation amplitudes --- for pristine and PEGylated virions. 

The difference of persistence lengths between M13 and Y21M phages is then sufficient to account, with $h=2.8L_p$,
for the magnitude and sign of the M13-PEG cholesteric pitch, \textit{as well} as for the nematic-like behavior (diverging pitch $P$) of the Y21M-PEG suspensions (Fig.~\ref{SupraHelix}). 
This mechanism of chirality transfer, based on long-wavelength helical deformations of the whole virus shape, relies mostly on excluded volume interactions, and may be shown using scaling arguments 
to depend solely on the ratio $\Phi/\Phi_{\rm iso}\equiv C_{virus}/C_{iso}$ for a fixed particle length $L_c$ (see Supplementary Section~V).

In light of the vanishingly-small values of $2\pi/P$ associated with the electrostatic model of ground-state M13 conformations (see Supplementary Section~III), let us postulate that the effects of the detailed chiral surface charge distribution on the cholesteric assembly of thermalized viruses may be neglected. We may then simply consider the viruses as uniformly-charged (helical) rods, whose liquid-crystalline behavior can be remapped to that of a hard-body system with a charge- and ionic-strength-dependent effective diameter $d_{\rm eff} > d$~\cite{
Tang1995}. 
Renormalizing the M13 virus concentration by the corresponding binodal value $C_{iso}$ at different ionic strengths (Extended Data Fig.~\ref{Y12M&M13rawData_pH8}), the inverse cholesteric pitch $P$ of both charged M13 and PEGylated M13-PEG 
is indeed found to collapse onto a unique \textit{master curve} (Fig.~\ref{MasterCurve}), quantitatively accounted for by the suprahelix model with $h=2.8L_p$.
Thus, contrary to the case of Y21M viruses where the local symmetry and details of the surface charge distribution matter, 
the existence of this master curve for M13 viruses proves 
the irrelevance of these local features for chirality propagation in the cholesteric phase, as quantitatively evidenced by the chiral potential of mean force (see Supplementary Section~VI
). 
Furthermore, it demonstrates that the origin of their chirality instead lies in long-wavelength helicoidal deformation modes, which are nearly not 
affected by PEGylation or changes in ionic conditions. 

\section{Conclusion and outlook}

We have extensively investigated the origin of chirality in liquid crystals of viruses --- a question that has largely eluded 
scientists for more than two decades~\cite{Grelet2003}. Through the fine tuning of experimental assembly conditions, combined with the development of quantitative atomistic models of the virus capsid, we are able to elucidate the mechanistic basis of chirality transfer 
for two distinct phages, M13 and Y21M, which, despite their high structural similarity, exhibit cholesteric phases of opposite handedness. The exquisite level of control over the molecular structure of the particles provided by these virus systems enables us to demonstrate that their surprising diversity of cholesteric behaviors stems from a subtle 
competition between thermal fluctuations, steric and electrostatic forces. For stiff Y21M virus strains, whose high bending rigidity largely suppresses conformational fluctuations away from a straight, rod-like backbone shape, we find that the cholesteric behavior may be quantitatively attributed to local electrostatic interactions, which are highly sensitive to both ionic content and the detailed atomic symmetries of the capsids. This conclusion qualitatively mirrors the findings of previous studies of various biopolymers including 
chitin~\cite{Narkevicius2019} and cellulose nanocrystals~\cite{Revol1998}. However, in contrast to these broadly-studied colloidal cholesteric systems, we report that the phase chirality of Y21M solutions decreases by reducing the strength of electrostatic interactions through the modulation of the pH and ionic environment. We attribute this behavior 
to the well-defined linear 
morphology of the virus backbone, whose molecular chirality originates primarily 
from the subtle helical distribution of surface charges around the symmetry axes of the capsid (Fig.~\ref{PyMol-Helix&ChargeDistrib}), and whose effects may be increasingly screened by reducing the range and magnitude of the associated electrostatic forces (see Supplementary Section VI). This ideal rod-like architecture differs from the 
size and shape polydispersity characterizing cholesteric assemblies of other bio-colloids such as chitin and cellulose, whose detailed atomistic structures remain challenging to resolve at the molecular level, as discussed further below. 
\par
Conversely, for the more flexible M13 variant, our results reveal that phase chirality instead proceeds from weak, fluctuation-induced supra-helical deformations of the virus backbone --- and is chiefly driven by steric, rather than electrostatic interactions. This conclusion mirrors recent findings in cholesteric systems of DNA origamis, and is evidenced by the collapse of the cholesteric pitches measured in various experimental conditions onto a unique master function of the rescaled concentration $C/C_{iso}$ (Fig.~\ref{MasterCurve}). In this context, the apparent unwinding of the cholesteric pitch with increasing ionic strengths at fixed (absolute) concentration $C$ may be quantitatively ascribed to the variations of the virion effective diameter $d_{\rm eff}$ (in the limit where the Debye screening length $\kappa^{-1} \ll L_c$), and is largely independent of the specific capsid surface charge patterns (see Supplementary Section V). Therefore, we emphasize that in order to rigorously assess the role of electrostatic interactions on chiral nematic ordering, one should carefully consider the variations of the cholesteric pitch at different pH and ionic conditions as a function of the reduced density $C/C_{iso}$, so as to accurately distinguish the potential contributions attributable to the chiral charge distribution from those arising from the generic effects of electrostatic forces on the nematic stability range. This task is likely arduous in cholesteric phases of fibrous biomaterials lacking a well-resolved, monodisperse 
morphology, for which the determination of $C_{iso}$ is necessarily ambiguous --- thus hindering the unequivocal experimental characterization 
of these distinct steric and electrostatic chirality transfer mechanisms in such cases. However, another experimental signature of the steric-based suprahelix model lies in the tightening of the cholesteric pitch with increasing particle contour lengths $L_c$
(see Supplementary Section~VII), which contradicts classical theoretical predictions for screw-like particles \cite{Odijk1987} and rod-shaped particles featuring helical surface charges~\cite{Kornyshev2002}. Interestingly, a similar behavior has been recently reported in fractionated cellulose~\cite{Honorato-Rios2020} as well as in amyloid fibril suspensions~\cite{Bagnani2019}, which suggests that such 
excluded-volume-driven modes of chirality propagation may find broader potential applications within a wider class of experimental colloidal and biological 
liquid crystalline systems. Overall, our 
findings emphasize how chirality transfer may  
arise from subtle chiral contributions of opposite handedness, which, depending on their magnitude and sign, can either act 
synergistically 
or competitively, therefore resulting in %
the diversity of chiral phase behaviors    
observed in 
nearly identical 
systems. 
\par
By combining 
both mesoscopic experimental measurements and theoretical predictions based on first-principle atomistic models, our work further provides a general methodological framework for the bottom-up, quantitative description of chirality transfer across length scales in cholesteric liquid crystals.
Understanding and controlling how chirality is expressed and transmitted in such helical superstructures 
may not only shed light on the various self-assembly processes and mechanisms leading to the large diversity of chiral liquid crystalline organizations, 
but also holds 
promise for the design of novel chiral materials with tailored optical, electronic, or biological functionalities \cite{Zhang2022,Sang2022,Kotov2022}. 

\section{Methods}

\textbf{Virus strains and capsid symmetries.} For this study, we employ two strains of filamentous bacteriophages referred to as M13 and Y21M \cite{Marvin1994,Marvin2014}. Both viruses belong to the Ff phage family, which includes the fd virus, and 
infect male (F$^+$) 
 \textit{Escherichia coli} (\textit{E. coli}).
 The phage capsid, protecting a single-stranded circular DNA core, is primarily comprised of about 3000 copies of 50-residue, identical
$\alpha$-helical subunits, called p8 proteins, assembled into an overlapping, interdigitated helical
structure.
 M13 and Y21M bear a high genetic similarity to the fd strain, 
implying that structural and biological properties are strongly conserved across these virions. 
In details, M13 and fd only differ by a single amino acid mutation at position 12 
of the major coat protein p8, whereby
 a negatively-charged aspartate (Asp) in fd
 is substituted by a neutral asparagine (Asn) in M13. 
This change in the solvent-accessible section of the protein modifies its overall charge, but does not structurally affect the symmetry of the two viruses. Indeed, both virions have been shown to have an identical capsid structure, as determined by X-ray fiber diffraction and NMR studies, deposited under the 1IFI model in the Protein Data Bank (PDB) \cite{Marvin1994,Marvin2006,Morag2015}.
Conversely, for Y21M, the single amino acid mutation occurs deeper within the capsid at position 21 in the coat proteins p8, where a tyrosine (Tyr) in M13 and fd is replaced by a methionine (Met) in Y21M. This substitution alters the symmetry of the phage capsid, and leads to 
a 5-fold rotation axis combined with an exact 2-fold helical axis (C$_5$S$_2$ symmetry), associated with a precise rotation angle of 36$^{{\rm o}}$ between two consecutive pentameric rings of p8 proteins --- as described in the 2C0W model of the PDB. This contrasts with the 1IFI model 
of M13, for which this angle is measured to be 33.23$^{{\rm o}}$ corresponding to a slight deviation from the two-fold screw symmetry \cite{Marvin2014}  
Another difference between the two strains concerns the rise 
between two pentameric subunits, varying from 1.60 to 1.67~nm depending on the model (Fig.~\ref{SchemePymol}) \cite{Marvin2006,Morag2015}. 
As a consequence of the spatial
arrangement of the major coat proteins, a main geometrical groove as deep as 1~nm 
\cite{Kishchenko1994} exists on both phage capsids, corresponding to a right-handed thread characterized by the angle $\varphi = 39.85^{{\rm o}}$ for M13 and $43.15^{{\rm o}}$ for Y21M (Figs. \ref{SchemePymol} and \ref{PyMol-Helix&ChargeDistrib}). 
A second right-handed thread can be identified on the M13 capsid, resulting from the lack of exact C$_5$S$_2$ symmetry in the IIFI model and characterized by the angle $\psi = 5.15^{{\rm o}}$, whereas this groove is achiral for Y21M (2C0W model) with $\psi = 0^{{\rm o}}$ (Fig. \ref{SchemePymol}).
These subtle structural differences between the two phages result in a 
large change of one of their physical properties, i.e. their stiffness, with a persistence length of $L_p=9.9~\mu$m for Y21M that has to be compared to $L_p=2.8~\mu$m for the semi-flexible M13 strain \cite{Barry2009}. 
Conversely, their contour lengths $L_c$ measured by transmission electron microscopy \cite{Pouget2011} are very close, with $L_c$=995 and 920~nm for M13 and Y21M respectively, and both viruses have the same diameter of about $d$=7~nm (Fig.~\ref{SchemePymol}).
The major coat proteins, p8, 
provide both viruses with a
helical charge distribution and a negative surface charge in physiological conditions, carried by ionic amino
acids exposed to the aqueous solvent. This net charge decreases as the buffer pH approaches the 
isoelectric point $pI_E$ (Fig.~\ref{PyMol-Helix&ChargeDistrib}), which is in the range 4.2-4.5 for Y21M and M13 \cite{Zimmermann86}.

\textbf{Virus preparation.} Both M13 and Y21M viruses are grown using the ER2738 strain as \textit{E. coli} host
bacteria and purified following standard biological protocols.
Virus PEGylation is performed
by covalent binding between coat protein amino groups and
N-hydroxysuccinimide ester-activated poly(ethylene glycol) (PEG)
of average molecular weight 21~kg.mol$^{-1}$ and radius of gyration $R_g \approx$~7~nm, 
as described with more details in reference \cite{Grelet2016}. PEGylation results in about 330 PEG chains per virus, and the associated phase behavior 
is shown to be driven by steric repulsion, i.e. to be independent of ionic strength \cite{Grelet2016}. Note that the 
isoelectric point and persistence length of PEGylated viruses are 
expected to be 
similar to that of unmodified viruses, 
as only a limited fraction of about 10\% of the coat proteins are bound with PEG \cite{Zan2016}. By neglecting the curvature of the virion capsid, this coverage with PEG polymers 
yields an 
upper-bound estimate for the surface
lateral pressure $\Pi \sim \SI{0.1}{~\milli\newton\per\meter}$ \cite{Marsh2001}, associated with a total effective force $F\propto \Pi/R_g \times d L_c \sim \SI{10}{~\pico \newton}$
exerted onto the capsid --- 
more than two orders of magnitude below the typical elastic stretching modulus reported for such filamentous viruses
\cite{Khalil2007}.
The virus concentration $C_{virus}$ at each dilution level is determined using spectrophotometry \cite{Barry2009} with an error bar of $\pm 2$~mg/mL.
For PEGylated viruses, the volume fraction $\Phi$ is calculated using an effective rod diameter $d_{\rm eff} = d + 4 R_g$ according to: 
$\Phi=\frac{C_{virus} N_A}{M_w}\times \frac{\pi}{4} L_{c}d_{\rm eff}^2$, with $M_W$ 
the virus molecular weight, and $N_A$ Avogadro's number \cite{Grelet2016}.
In order to study the pH dependence of the cholesteric pitch, different biological buffers are used covering basic to acidic conditions: 2-(Cyclohexylamino)ethanesulfonic acid (CHES, $pK_a$~9.3); (Tris(hydroxymethyl)aminomethane (TRIS, $pK_a$~8.2); 4-Morpholineethanesulfonic acid (MES, $pK_a$~6.1); pyridine ($pK_a$~5.2); propianic acid ($pK_a$~4.9); acetic acid ($pK_a$~4.7); chloroacetic acid ($pK_a$~2.9).
The buffering agent is introduced with an analytical concentration of 20~mM, and pH and ionic strength $I_s$ are adjusted with tunable amounts of NaOH (or HCl) and NaCl, respectively. After extensive dialysis, virus suspensions of different dilutions are prepared in previously cleaned (successive rinsing with acetone, isopropanol, distilled water  followed by 30 min of UV-ozone treatment (Harrick Plasma)) quartz capillary tubes of diameter around 1.5~mm for polarizing microscopy observations, and between cover slip and glass slide with parafilm spacer of about 100~$\mu m$ for samples used for handedness determination by fluorescence microscopy. In the latter case, samples are doped at a fraction of 1:10$^5$ with red- or green-labeled viruses, grafted with Dylight550-NHS ester (ThermoFischer) and Alexa488-NHS-ester (ThermoFischer), respectively. 

\textbf{Optical microscopy experiments.} Epifluorescence images are obtained using an inverted optical microscope (IX71,
Olympus) equipped with a 100× oil-immersion objective (NA 1.4, UPLSAPO), a
piezo device for objective z-positioning (P-721 PIFOC Piezo Flexure Objective
Scanner, PI) operated by computer interface software (Meta-Morph, Molecular
Devices), a LED light engine (LedHUB, Omicron), and a fluorescence
imaging camera (NEO sCMOS, Andor Technology). The mirror symmetry of the whole optical setup is checked before each observation. The kinetics of establishing the chiral nematic phase is shown to be sample-dependent: a few days to a few weeks of equilibration are applied to get homogeneous fingerprint textures (Inset of Fig.~\ref{Y21M&M13vsElectroModel} and Extended Data Fig.~\ref{Texture}f), resulting from the virion planar anchoring at the capillary walls. Cholesteric pitch measurements are carried out at low magnification (5× LMPlanFl objective, NA 0.13) with a polarizing microscope (BX-51, Olympus) equipped with a JAI-CV-M7+ color camera.
The cholesteric pitch determination obtained from fingerprint textures is an average value of 
10 to 20 measurements and error bars correspond to the 
dispersion of the experimental values. The nematic phase is ascribed to samples for which no helical pitch is observed over the millimeter range, corresponding to the capillary diameter. 

\textbf{Numerical methods.} The molecular structures of the whole capsids of M13 and Y21M viruses are respectively reconstructed from the atomic models 1IFI and 2C0W deposited in the PDB (see above)~\cite{Marvin2006,Morag2015}. Potential energies are parametrized using the GROMOS 53A6 force field~\cite{oostenbrink2004} to explicitly account for van der Waals and excluded-volume interactions between each pairs of atoms within the capsids. Screened electrostatic contributions are described by the generalized reaction field method~\cite{Tironi1995}, corresponding to a computationally-efficient treatment of long-ranged electrostatic forces based on the Kirkwood-Onsager continuum theory of dielectric polarization~\cite{Tironi1995}. In particular, this approach allows for the use of a finite truncation cutoff (see below), combined with an implicit representation of the ionic environment in terms of the solvent ionic strength, temperature and dielectric permittivity --- and has been shown to accurately capture the thermodynamics of a wide variety of protein condensates~\cite{oostenbrink2004}. Molecular charge distributions are determined using the PROPKA plugin~\cite{
Olsson2011propka} of the pdb2pqr pipeline~\cite{Dolinsky2007}. 
For computational tractability, a cutoff radius $r_{\rm cut}=3.5$~nm is applied in the calculation of all electrostatic energies. This approximation is expected to hold in the limit where the Debye screening length $\kappa^{-1}$ is such that $\kappa^{-1} \ll r_{\rm cut}$, which is typically valid for ionic strengths $I_s \ge 100~{\rm mM}$. In this context, the effective steric contribution is computed by evaluating the full force field in the limit of high salt concentrations ($I_s = 1~{\rm M}$). 

For the computation of cholesteric pitches, the system free energy $\mathscr{F}$ is derived for both the electrostatic and suprahelix models at the second virial level as a functional of the full intermolecular pairwise potential $U_{\rm inter}$, based on a perturbative expansion of the 
Onsager expression $\mathscr{F}_0$ for the uniform, untwisted nematic state~\cite{Onsager1949}. The optimal angular arrangement of the particles about the local director, which generally depends on the detailed virus structure and thermodynamic state, is determined by functional minimization of $\mathscr{F}_0$ at fixed concentration~\cite{Tortora2020}. The corresponding equilibrium pitch $P$ and the twist elastic constant $K_{22}$ are obtained by subsequent minimization of the full free energy $\mathscr{F}$ based on the computed local orientational distribution (Supplementary Section~I), and may be formulated in terms of a hierarchy of generalized virial integrals involving $U_{\rm inter}$~\cite{Tortora2020}. Such integrals are evaluated via high-performance Monte-Carlo sampling techniques as described elsewhere~\cite{Tortora2017b}. Importantly, this framework enables us to accurately infer the most-favorable, large-scale (micro- to millimeter range) cholesteric structure from the atomistic details of the different virus models, with a level of precision tunable through the statistical resolution of the stochastic integration scheme~\cite{Tortora2020}. Accordingly, error bars are estimated as the standard error of the computed pitches across $\mathcal{O}(10)$ independent Monte-Carlo runs, using a number $\mathcal{O}(10^{14})$ of integration steps. Binodal points are calculated by equating chemical potentials and osmotic pressures in the isotropic and cholesteric phases, and solving the resulting coupled coexistence equations numerically~\cite{Tortora2020}. 

\medbreak
\textbf{Acknowledgments} We thank H.~Anop for the data of Extended Data Figure 5, 
and A.~Pope for help with Y21M sample preparation. We also acknowledge access to computing resources provided by 
the P\^ole Scientifique de Mod\'elisation Num\'erique of the ENS de Lyon.
\medbreak
\textbf{Author contributions} E.G.~conceptualized 
the study, instigated the project, performed the experiments and wrote the manuscript with contributions from M.M.C.T. 
M.M.C.T.~implemented the numerical methods and carried out the calculations. Both authors developed the models, analyzed the results, wrote the Supplementary Information, revised and edited the manuscript. 
\medbreak
\textbf{Competing interests} The authors declare no competing interests. 
\medbreak
\textbf{Additional information} Supplementary information available. Correspondence and requests for materials and data should be addressed to E.G. Queries regarding numerical details and computational data should be directed to M.M.C.T.
\medbreak
\textbf{Data and code availability} Data generated or analyzed during this study are included in the Article and the Supplementary Information. 

Numerical codes used for molecular structure preparation and density functional calculations are available on \href{https://github.com/mtortora/chiralDFT}{GitHub}. 

\bibliographystyle{unsrt}
\bibliography{Rods}	

\begin{thebibliography}{10}

\bibitem{Liu2015}
Minghua Liu, Li~Zhang, and Tianyu Wang.
\newblock Supramolecular chirality in self-assembled systems.
\newblock {\em Chemical Reviews}, 115(15):7304--7397, Aug 2015.

\bibitem{Morrow2017}
Sarah~M. Morrow, Andrew~J. Bissette, and Stephen~P. Fletcher.
\newblock Transmission of chirality through space and across length scales.
\newblock {\em Nature Nanotechnology}, 12(5):410--419, May 2017.

\bibitem{Nemati2018}
Ahlam Nemati, Sasan Shadpour, Lara Querciagrossa, Lin Li, Taizo Mori, Min Gao,
  Claudio Zannoni, and Torsten Hegmann.
\newblock Chirality amplification by desymmetrization of chiral ligand-capped
  nanoparticles to nanorods quantified in soft condensed matter.
\newblock {\em Nature Communications}, 9(1):3908, Sep 2018.

\bibitem{Nemati2022}
Ahlam Nemati, Lara Querciagrossa, Corinne Callison, Sasan Shadpour, Diana
  P.~Nunes Gonçalves, Taizo Mori, Ximin Cui, Ruoqi Ai, Jianfang Wang, Claudio
  Zannoni, and Torsten Hegmann.
\newblock Effects of shape and solute-solvent compatibility on the efficacy of
  chirality transfer: Nanoshapes in nematics.
\newblock {\em Science Advances}, 8(4):eabl4385, 2022.

\bibitem{Zhang2022}
Xuan Zhang, Yiyi Xu, Cristian Valenzuela, Xinfang Zhang, Ling Wang, Wei Feng,
  and Quan Li.
\newblock Liquid crystal-templated chiral nanomaterials: from chiral plasmonics
  to circularly polarized luminescence.
\newblock {\em Light: Science {\&} Applications}, 11(1):223, Jul 2022.

\bibitem{Sang2022}
Yutao Sang and Minghua Liu.
\newblock Hierarchical self-assembly into chiral nanostructures.
\newblock {\em Chem. Sci.}, 13:633--656, 2022.

\bibitem{Kotov2022}
Nicholas~A. Kotov, Luis~M. Liz-Marzán, and Qiangbin Wang.
\newblock Chiral nanomaterials: evolving rapidly from concepts to applications.
\newblock {\em Mater. Adv.}, 3:3677--3679, 2022.

\bibitem{Mitov2012}
Michel Mitov.
\newblock Cholesteric liquid crystals with a broad light reflection band.
\newblock {\em Advanced Materials}, 24(47):6260--6276, 2012.

\bibitem{Geng2022}
Yong Geng, Rijeesh Kizhakidathazhath, and Jan P.~F. Lagerwall.
\newblock Robust cholesteric liquid crystal elastomer fibres for mechanochromic
  textiles.
\newblock {\em Nature Materials}, 21(12):1441--1447, Dec 2022.

\bibitem{Bisoyi2022}
Hari~Krishna Bisoyi and Quan Li.
\newblock Liquid crystals: Versatile self-organized smart soft materials.
\newblock {\em Chemical Reviews}, 122(5):4887--4926, 2022.

\bibitem{Mitov2017}
Michel Mitov.
\newblock Cholesteric liquid crystals in living matter.
\newblock {\em Soft Matter}, 13(23):4176--4209, 2017.

\bibitem{Reinitzer1888}
Friedrich Reinitzer.
\newblock Beitr{\"a}ge zur kenntniss des cholesterins.
\newblock {\em Monatshefte f{\"u}r Chemie}, 9:421--441, 1888.

\bibitem{Livolant1996}
Fran{\c{c}}oise Livolant and Am{\'e}lie Leforestier.
\newblock Condensed phases of {DNA}: structures and phase transitions.
\newblock {\em Prog. Polym. Sci.}, 21(6):1115--1164, 1996.

\bibitem{Zanchetta2010}
G.~Zanchetta, F.~Giavazzi, M.~Nakata, M.~Buscaglia, R.~Cerbino, N.~A. Clark,
  and T.~Bellini.
\newblock Right-handed double-helix ultrashort {DNA} yields chiral nematic
  phases with both right- and left-handed director twist.
\newblock {\em Proc. Natl. Acad. Sci. USA}, 107:17497--17502, 2010.

\bibitem{Siavashpouri2017}
M.~Siavashpouri, C.~H. Wachauf, M.~J. Zakhary, F.~Praetorius, H.~Dietz, and
  Z.~Dogic.
\newblock Molecular engineering of chiral colloidal liquid crystals using {DNA}
  origami.
\newblock {\em Nat. Mater.}, 16:849--856, 2017.

\bibitem{Dogic2000}
Zvonimir Dogic and Seth Fraden.
\newblock Cholesteric phase in virus suspensions.
\newblock {\em Langmuir}, 16(20):7820--7824, 2000.

\bibitem{Grelet2003}
Eric Grelet and Seth Fraden.
\newblock What is the origin of chirality in the cholesteric phase of virus
  suspensions?
\newblock {\em Phys. Rev. Lett.}, 90(19):198302, 2003.

\bibitem{Tombolato2006}
Fabio Tombolato, Alberta Ferrarini, and Eric Grelet.
\newblock Chiral nematic phase of suspensions of rodlike viruses: left-handed
  phase helicity from a right-handed molecular helix.
\newblock {\em Phys. Rev. Lett.}, 96(25):258302, 2006.

\bibitem{Bagnani2019}
Massimo Bagnani, Gustav Nystr{\"o}m, Cristiano De~Michele, and Raffaele
  Mezzenga.
\newblock Amyloid fibrils length controls shape and structure of nematic and
  cholesteric tactoids.
\newblock {\em ACS Nano}, 13(1):591--600, 01 2019.

\bibitem{Belamie2004}
E.~Belamie, P.~Davidson, and M.~M. Giraud-Guille.
\newblock Structure and chirality of the nematic phase in $\alpha$-chitin
  suspensions.
\newblock {\em The Journal of Physical Chemistry B}, 108(39):14991--15000, Sep
  2004.

\bibitem{Araki2001b}
Jun Araki and Shigenori Kuga.
\newblock Effect of trace electrolyte on liquid crystal type of cellulose
  microcrystals.
\newblock {\em Langmuir}, 17(15):4493--4496, Jul 2001.

\bibitem{Usov2015}
Ivan Usov, Gustav Nystr{\"o}m, Jozef Adamcik, Stephan Handschin, Christina
  Sch{\"u}tz, Andreas Fall, Lennart Bergstr{\"o}m, and Raffaele Mezzenga.
\newblock Understanding nanocellulose chirality and structure--properties
  relationship at the single fibril level.
\newblock {\em Nature Communications}, 6(1):7564, Jun 2015.

\bibitem{Honorato-Rios2020}
Camila Honorato-Rios and Jan P.~F. Lagerwall.
\newblock Interrogating helical nanorod self-assembly with fractionated
  cellulose nanocrystal suspensions.
\newblock {\em Communications Materials}, 1(1):69, Sep 2020.

\bibitem{Parton2022}
Thomas~G. Parton, Richard~M. Parker, Gea~T. van~de Kerkhof, Aurimas
  Narkevicius, Johannes~S. Haataja, Bruno Frka-Petesic, and Silvia Vignolini.
\newblock Chiral self-assembly of cellulose nanocrystals is driven by
  crystallite bundles.
\newblock {\em Nature Communications}, 13(1):2657, May 2022.

\bibitem{Straley1976}
Joseph~P. Straley.
\newblock Theory of piezoelectricity in nematic liquid crystals, and of the
  cholesteric ordering.
\newblock {\em Phys. Rev. A}, 14(5):1835, 1976.

\bibitem{Harris1999}
A~Brooks Harris, Randall~D Kamien, and Thomas~C Lubensky.
\newblock Molecular chirality and chiral parameters.
\newblock {\em Rev. Mod. Phys.}, 71(5):1745, 1999.

\bibitem{Osipov1988}
M.~A. Osipov.
\newblock Theory for cholesteric ordering in lyotropic liquid crystals.
\newblock {\em Il Nuovo Cimento D}, 10(11):1249--1262, Nov 1988.

\bibitem{Cherstvy2008}
A.~G. Cherstvy.
\newblock {DNA} cholesteric phases: The role of {DNA} molecular chirality and
  {DNA} electrostatic interactions.
\newblock {\em J. Phys. Chem. B}, 142:12585--12595, 2008.

\bibitem{Dussi2016}
Simone Dussi and Marjolein Dijkstra.
\newblock Entropy-driven formation of chiral nematic phases by computer
  simulations.
\newblock {\em Nature Communications}, 7(1):11175, Apr 2016.

\bibitem{Tortora2020}
Maxime M.~C. Tortora, Garima Mishra, Domen Prešern, and Jonathan P.~K. Doye.
\newblock Chiral shape fluctuations and the origin of chirality in cholesteric
  phases of dna origamis.
\newblock {\em Science Advances}, 6(31):eaaw8331, 2020.

\bibitem{Dogic&Fraden}
Zvonimir Dogic and Seth Fraden.
\newblock {\em Soft Matter Vol. 2: Complex Colloidal Suspensions, edited by G.
  Gompper and M. Schick}.
\newblock Wiley-VCH, Weinheim, 2006.

\bibitem{Smith1997}
George~P. Smith and Valery~A. Petrenko.
\newblock Phage display.
\newblock {\em Chemical Reviews}, 97(2):391--410, Apr 1997.

\bibitem{Marvin2006}
D.~A. Marvin, L.~C. Welsh, M.~F. Symmons, W.~R.~P. Scott, and S.~K. Straus.
\newblock Molecular structure of fd (f1, {M}13) filamentous bacteriophage
  refined with respect to {X}-ray fibre diffraction and solid-state {NMR} data
  supports specific models of phage assembly at the bacterial membrane.
\newblock {\em Journal of molecular biology}, 355:294--309, 02 2006.

\bibitem{Lee2009}
Yun~Jung Lee, Hyunjung Yi, Woo-Jae Kim, Kisuk Kang, Dong~Soo Yun, Michael~S.
  Strano, Gerbrand Ceder, and Angela~M. Belcher.
\newblock Fabricating genetically engineered high-power lithium-ion batteries
  using multiple virus genes.
\newblock {\em Science}, 324(5930):1051--1055, 2009.

\bibitem{Marvin2014}
D.~A. Marvin, M.~F. Symmons, and S.~K. Straus.
\newblock Structure and assembly of filamentous bacteriophages.
\newblock {\em Progress in Biophysics and Molecular Biology}, 114:80--122,
  2014.

\bibitem{Gibaud2012}
T.~Gibaud, E.~Barry, M.~Zakhary, M.~Henglin, A.~Ward, Y.~Yang, C.~Berciu,
  R.~Oldenbourg, M.~Hagan, D.~Nicastro, R.~Meyer, and Z.~Dogic.
\newblock Self-assembly through chiral control of interfacial tension.
\newblock {\em Nature}, 481:348, 2012.

\bibitem{Grelet2014}
Eric Grelet.
\newblock Hard-rod behavior in dense mesophases of semiflexible and rigid
  charged viruses.
\newblock {\em Phys. Rev. X}, 4:021053, 2014.

\bibitem{Willis2008}
Bert Willis, Lisa~M. Eubanks, Malcom~R. Wood, Kim~D. Janda, Tobin~J. Dickerson,
  and Richard~A. Lerner.
\newblock Biologically templated organic polymers with nanoscale order.
\newblock {\em Proc. Natl. Acad. Sci. USA}, 105:1416--1419, 2008.

\bibitem{Chung2011}
Woo-Jae Chung, Jin-Woo Oh, Kyungwon Kwak, Byung~Yang Lee, Joel Meyer, Eddie
  Wang, Alexander Hexemer, and Seung-Wuk Lee.
\newblock Biomimetic self-templating supramolecular structures.
\newblock {\em Nature}, 478(7369):364--368, Oct 2011.

\bibitem{Barry2009}
Edward Barry and Zvonimir Dogic.
\newblock A model liquid crystalline system based on rodlike viruses with
  variable chirality and persistence length.
\newblock {\em Soft Matter}, 5:2563--2570, 2009.

\bibitem{Jurrus2018}
Elizabeth Jurrus, Dave Engel, Keith Star, Kyle Monson, Juan Brandi, Lisa~E.
  Felberg, David~H. Brookes, Leighton Wilson, Jiahui Chen, Karina Liles, Minju
  Chun, Peter Li, David~W. Gohara, Todd Dolinsky, Robert Konecny, David~R.
  Koes, Jens~Erik Nielsen, Teresa Head-Gordon, Weihua Geng, Robert Krasny,
  Guo-Wei Wei, Michael~J. Holst, J.~Andrew McCammon, and Nathan~A. Baker.
\newblock Improvements to the {APBS} biomolecular solvation software suite.
\newblock {\em Protein Sci.}, 27(1):112--128, 2018.

\bibitem{Frezza2014}
Elisa Frezza, Alberta Ferrarini, Hima Bindu~Kolli, Achille Giacometti, and
  Giorgio Cinacchi.
\newblock Left or right cholesterics? a matter of helix handedness and
  curliness.
\newblock {\em Phys. Chem. Chem. Phys.}, 16:16225--16232, 2014.

\bibitem{Dussi2015}
Simone Dussi, Simone Belli, Ren{\'e} van Roij, and Marjolein Dijkstra.
\newblock Cholesterics of colloidal helices: Predicting the macroscopic pitch
  from the particle shape and thermodynamic state.
\newblock {\em J. Chem. Phys.}, 142(7):074905, 2015.

\bibitem{Kornyshev2002}
A.~A. Kornyshev, S.~Leikin, and S.~V. Malinin.
\newblock Chiral electrostatic interaction and cholesteric liquid crystals of
  {DNA}.
\newblock {\em Eur. Phys. J. E}, 7:83--93, 2002.

\bibitem{Wensink2009}
Henricus~H. Wensink and G.~Jackson.
\newblock Generalized van der waals theory for the twist elastic modulus and
  helical pitch of cholesterics.
\newblock {\em J. Chem. Phys.}, 130(23):234911, 2009.

\bibitem{Onsager1949}
Lars Onsager.
\newblock The effects of shape on the interaction of colloidal particles.
\newblock {\em Ann. N.Y. Acad. Sci.}, 51(4):627--659, 1949.

\bibitem{Zhang2014}
C.~Zhang, N.~Diorio, O.~D. Lavrentovich, and A.~J{\'a}kli.
\newblock Helical nanofilaments of bent-core liquid crystals with a second
  twist.
\newblock {\em Nature Communications}, 5(1):3302, Feb 2014.

\bibitem{oostenbrink2004}
Chris Oostenbrink, Alessandra Villa, Alan~E. Mark, and Wilfred~F.
  Van~Gunsteren.
\newblock A biomolecular force field based on the free enthalpy of hydration
  and solvation: The gromos force-field parameter sets 53a5 and 53a6.
\newblock {\em J. Comput. Chem.}, 25(13):1656--1676, 2004.

\bibitem{Grelet2016}
Eric Grelet and Richa Rana.
\newblock From soft to hard rod behavior in liquid crystalline suspensions of
  sterically stabilized colloidal filamentous particles.
\newblock {\em Soft Matter}, 12:4621, 2016.

\bibitem{Odijk1986}
Theo Odijk.
\newblock Theory of lyotropic polymer liquid crystals.
\newblock {\em Macromolecules}, 19(9):2313--2329, 1986.

\bibitem{Chen1993}
Zheng~Yu Chen.
\newblock Nematic ordering in semiflexible polymer chains.
\newblock {\em Macromolecules}, 26(13):3419--3423, 1993.

\bibitem{Tang1995}
Jianxin Tang and Seth Fraden.
\newblock Isotropic-cholesteric phase transition in colloidal suspensions of
  filamentous bacteriophage fd.
\newblock {\em Liquid Crystals}, 19(4):459--467, 1995.

\bibitem{Narkevicius2019}
Aurimas Narkevicius, Lisa~M. Steiner, Richard~M. Parker, Yu~Ogawa, Bruno
  Frka-Petesic, and Silvia Vignolini.
\newblock Controlling the self-assembly behavior of aqueous chitin nanocrystal
  suspensions.
\newblock {\em Biomacromolecules}, 20(7):2830--2838, 07 2019.

\bibitem{Revol1998}
J-F Revol, Louis Godbout, and Derek~G Gray.
\newblock Solid self-assembled films of cellulose with chiral nematic order and
  optically variable properties.
\newblock {\em J. Pulp Paper Sci.}, 24(5):146--149, 1998.

\bibitem{Odijk1987}
Theo Odijk.
\newblock Pitch of a polymer cholesteric.
\newblock {\em J. Phys. Chem.}, 91(23):6060--6062, 1987.

\bibitem{Marvin1994}
D.~A. Marvin, R.~D. Hale, C.~Nave, and M.~Helmer~Citterich.
\newblock Molecular models and structural comparisons of native and mutant
  class {I} filamentous bacteriophages.
\newblock {\em Journal of molecular biology}, 235:260--286, 02 1994.

\bibitem{Morag2015}
Omry Morag, Nikolaos~G. Sgourakis, David Baker, and Amir Goldbourt.
\newblock The \uppercase{NMR-R}osetta capsid model of {M}13 bacteriophage
  reveals a quadrupled hydrophobic packing epitope.
\newblock {\em Proc. Natl. Acad. Sci. USA}, 112(4):971--976, 2015.

\bibitem{Kishchenko1994}
Gregory Kishchenko, Hoshang Batliwala, and Lee Makowski.
\newblock Structure of a foreign peptide displayed on the surface of
  bacteriophage {M}13.
\newblock {\em J. Mol. Biol.}, 241:208--213, 1994.

\bibitem{Pouget2011}
E.~Pouget, E.~Grelet, and M.~P. Lettinga.
\newblock Dynamics in the smectic phase of stiff viral rods.
\newblock {\em Phys Rev E Stat Nonlin Soft Matter Phys}, 84:041704, 2011.

\bibitem{Zimmermann86}
K.~Zimmermann, H.~Hagedorn, C.~Chr. Heucks, M.~Hinrichsen, and H.~Ludwig.
\newblock The ionic properties of the filamentous bacteriophages pfl and fd.
\newblock {\em J. Biol. Chem.}, 261:1653--1655, 1986.

\bibitem{Zan2016}
Tingting Zan, Fengchi Wu, Xiaodong Pei, Shaoyi Jia, Ran Zhang, Songhai Wu,
  Zhongwei Niu, and Zhenkun Zhang.
\newblock Into the polymer brush regime through the “grafting-to” method:
  densely polymer-grafted rodlike viruses with an unusual nematic liquid
  crystal behavior.
\newblock {\em Soft Matter}, 12:798--805, 2016.

\bibitem{Marsh2001}
Derek Marsh.
\newblock Elastic constants of polymer-grafted lipid membranes.
\newblock {\em Biophysical Journal}, 81(4):2154--2162, 2001.

\bibitem{Khalil2007}
Ahmad~S. Khalil, Jorge~M. Ferrer, Ricardo~R. Brau, Stephen~T. Kottmann,
  Christopher~J. Noren, Matthew~J. Lang, and Angela~M. Belcher.
\newblock Single {M13} bacteriophage tethering and stretching.
\newblock {\em Proceedings of the National Academy of Sciences},
  104(12):4892--4897, 2007.

\bibitem{Tironi1995}
Ilario~G. Tironi, René Sperb, Paul~E. Smith, and Wilfred~F. van Gunsteren.
\newblock A generalized reaction field method for molecular dynamics
  simulations.
\newblock {\em J. Chem. Phys.}, 102(13):5451--5459, 1995.

\bibitem{Olsson2011propka}
Mats H.~M. Olsson, Chresten~R. S{\o}ndergaard, Michal Rostkowski, and Jan~H.
  Jensen.
\newblock {PROPKA3}: Consistent treatment of internal and surface residues in
  empirical p{K}a predictions.
\newblock {\em J. Chem. Theory Comput.}, 7(2):525--537, 02 2011.

\bibitem{Dolinsky2007}
Todd~J. Dolinsky, Paul Czodrowski, Hui Li, Jens~E. Nielsen, Jan~H. Jensen,
  Gerhard Klebe, and Nathan~A. Baker.
\newblock Pdb2pqr: expanding and upgrading automated preparation of
  biomolecular structures for molecular simulations.
\newblock {\em Nucleic Acids Res.}, 35($suppl_2$):W522--W525, 07 2007.

\bibitem{Tortora2017b}
Maxime M.~C. Tortora and Jonathan P.~K. Doye.
\newblock Hierarchical bounding structures for efficient virial computations:
  Towards a realistic molecular description of cholesterics.
\newblock {\em J. Chem. Phys.}, 147(22):224504, 2017.

\end{thebibliography}

\clearpage
\setcounter{figure}{0}
\renewcommand{\figurename}{
Extended Data Fig.}

\begin{figure}
	\includegraphics[width=0.7\columnwidth]{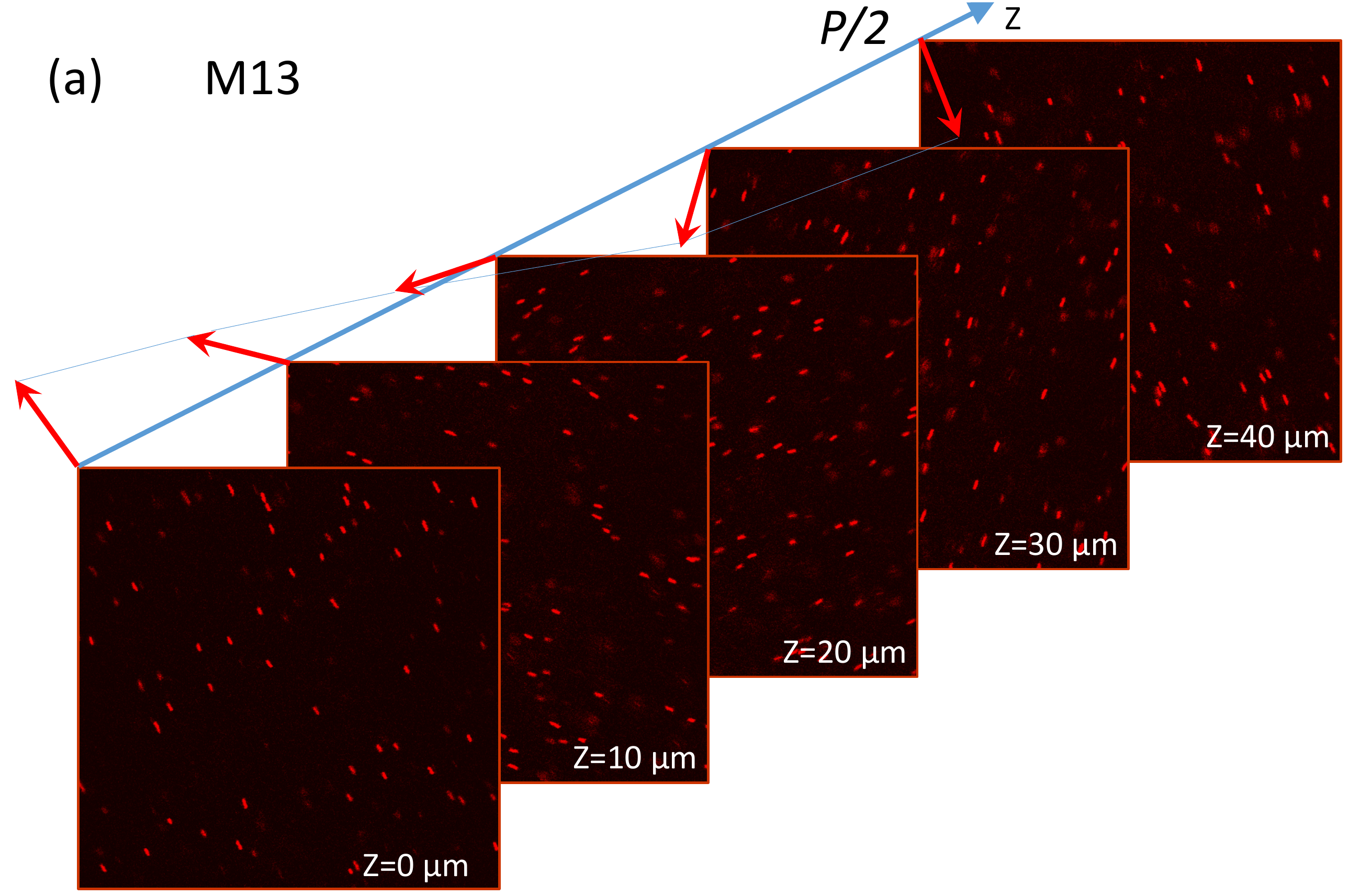}
	\includegraphics[width=0.7\columnwidth]{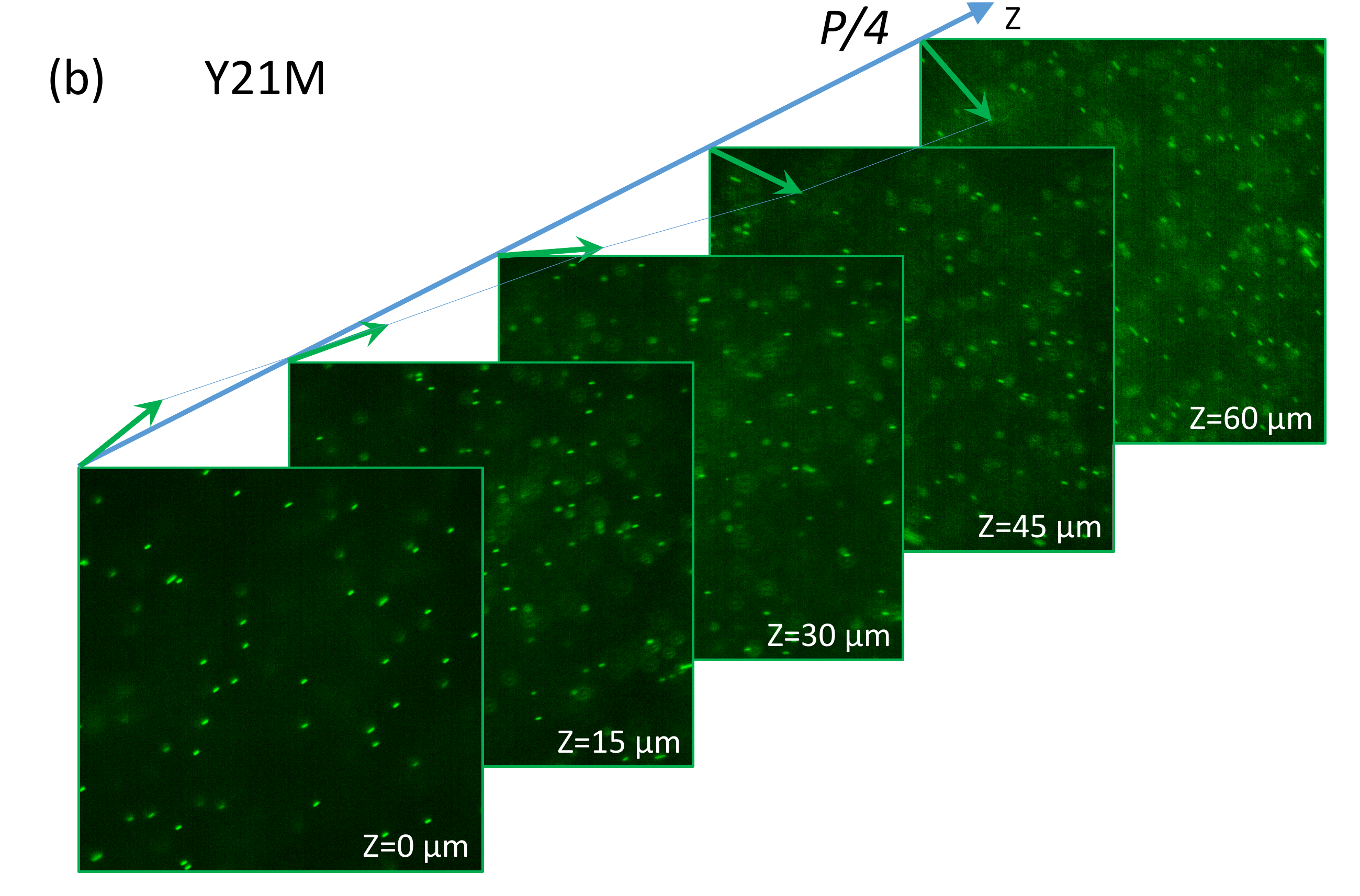}
	\caption{\textbf{Opposite handedness of the cholesteric helix for M13 and Y21M strains}, as determined by fluorescence microscopy in (a) PEGylated M13 suspension (pH 8.2, $I_S$=110~mM) and 
		(b) Y21M suspension (pH 8.2, $I_S$=60~mM). 
		A small fraction (1:10$^5$) of viruses are labelled with 
  red or green fluorescent tags to 
  indicate the orientation of the nematic director in each focal plane, as shown by arrows. Their rotation through the sample thickness Z reveals the handedness of the cholesteric helicity, which is found to be left-handed for 
  M13 and M13-PEG and right-handed for Y21M strain. The periodicity of the cholesteric helix, or cholesteric pitch $P$, is also indicated for both virion strains, and its value is positive (negative) for right (left) handedness. 
Each image has a size of 50~$\mu$m x 50~$\mu$m. }
		\label{Handedness}
\end{figure}

\begin{figure}
	\includegraphics[width=0.9\columnwidth]{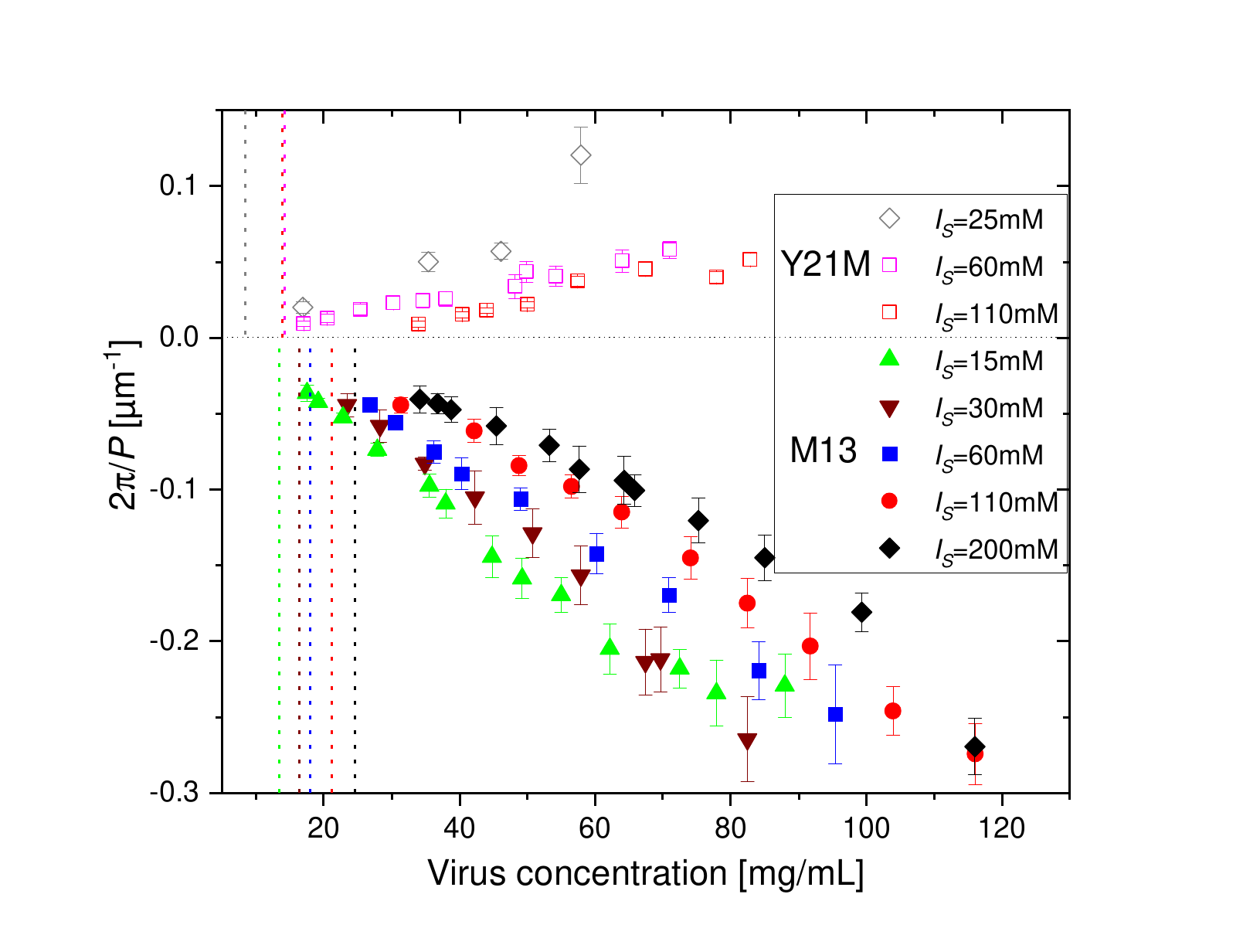}
	\caption{\textbf{Electrostatic dependence of the 
    cholesteric pitch $P$}, measured for Y21M (open symbols) and M13 (full symbols) for different ionic strengths $I_S$ at fixed pH~8 as a function of the respective virus concentration. The
    data of Y21M phages at pH~8 and $I_S=110$~mM are taken from  Ref.~\cite{Barry2009}. 
    For both virions, $\vert P \vert$ increases 
    with increasing ionic strength, i.e., with increasing the screening of electrostatic interactions. For each data set, the 
    binodal concentrations of the isotropic-to-cholesteric transition corresponding to the stability limit of the isotropic phase, $C_{iso}$, is shown 
    by a dotted line whose color corresponds to the associated color of the symbols.}
	\label{Y12M&M13rawData_pH8}
\end{figure}

\begin{figure}
	\includegraphics[width=0.75\columnwidth]{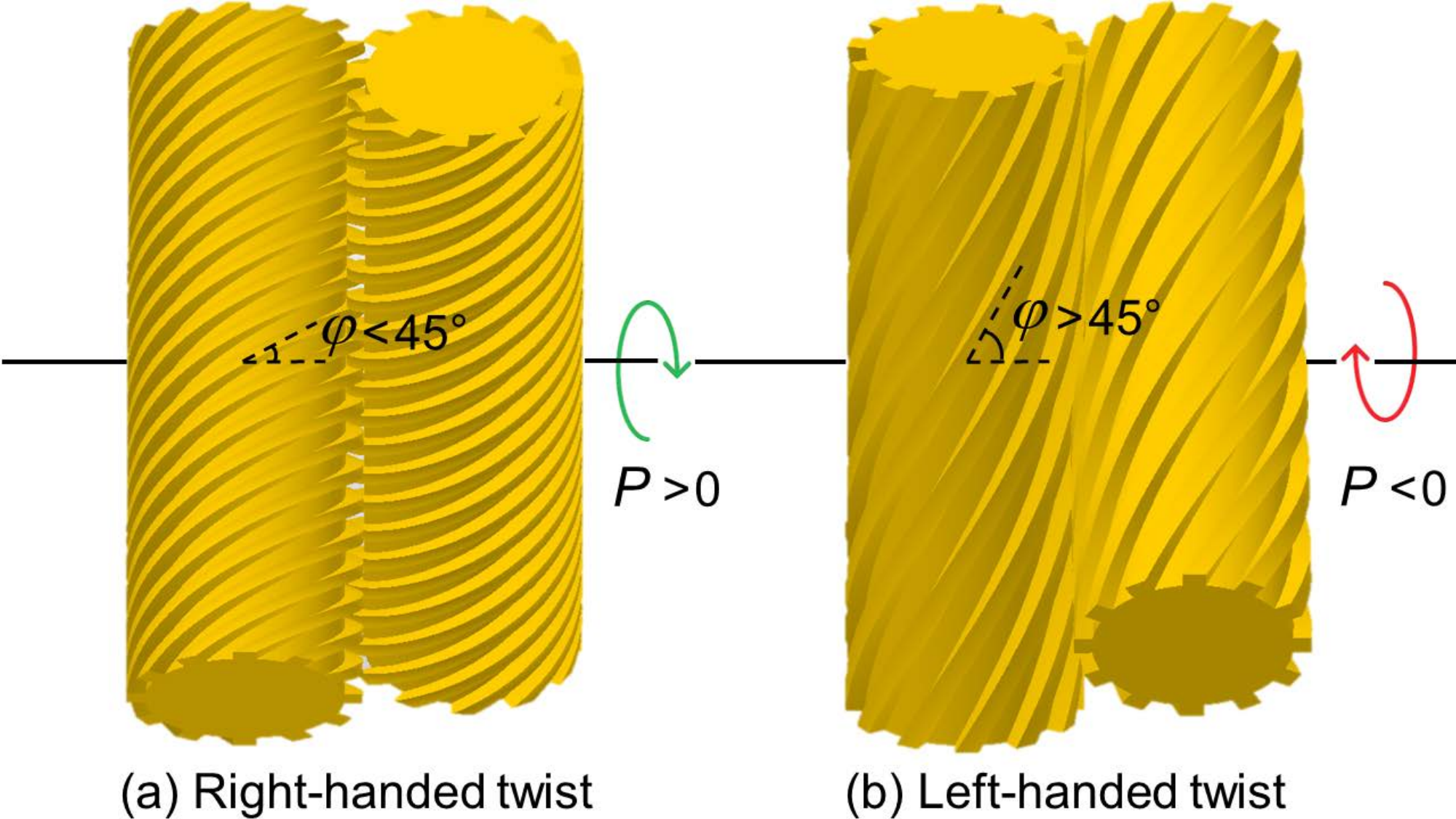}
	\caption{\textbf{Inversion of the twist handedness between right-handed screws of varying thread angle $\varphi$}. 
		The helical twist resulting from the close packing of two \textit{right}-handed (i.e. $0 < \varphi < +90^{{\rm o}}$) screws
		leads (a) to a \textit{right}-handed twist (and therefore a right-handed cholesteric pitch $P>0$) of angle $2\varphi>0$ when $\varphi 
		<45^{{\rm o}}$ and (b) to a \textit{left}-handed twist ($P<0$) of angle $-(180^{{\rm o}}-2\varphi)<0$ for $\varphi 
		>45^{{\rm o}}$. 
	} 
	\label{screws}
\end{figure}

\begin{figure}
	\includegraphics[width=0.7\columnwidth]{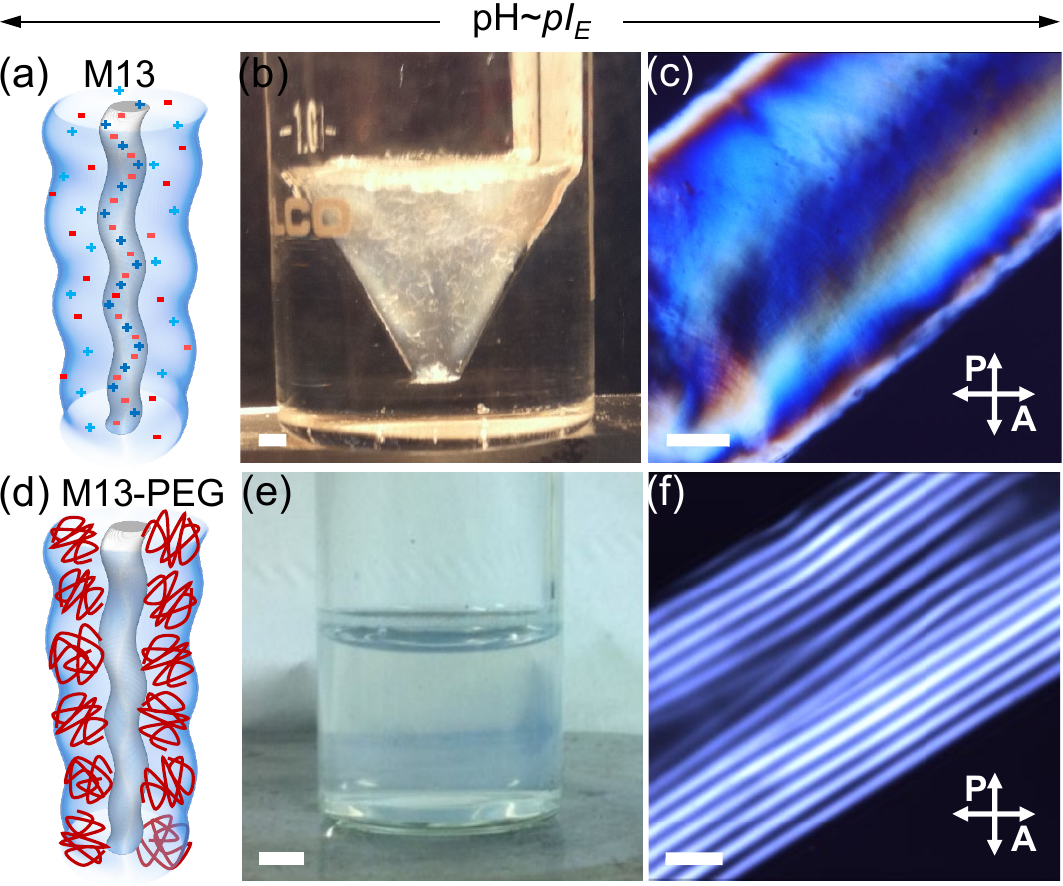}
	\caption{\textbf{Phase behavior of semi-flexible M13 (a)-(c)  
			 and PEGylated M13-PEG (d)-(f)
		virus suspensions at pH close the isoelectric point, 
		$pI_E$}. (a) and (d): Schematic representation of the filamentous viruses, whose colloidal stability stems from either (a) electrostatic or (d) steric repulsion. (b) and (e): Macroscopic observation under white light of the virion suspensions: while aggregates are observed in raw M13 virus dispersions at  pH$\simeq pI_E$, the colloidal stability is preserved in the M13-PEG system. Scale bar: 2~mm. (c) and (f): Polarized optical microscopy images displaying  
		a nematic-like birefingent texture with fibrillar moieties for raw M13 viruses
		 (c) and the characteristic fingerprint texture of the cholesteric phase for PEGylated particles (f). Scale bar: 200~$\mu$m. } 
	\label{Texture}
\end{figure}

\begin{figure}
	\includegraphics[width=0.75\columnwidth]{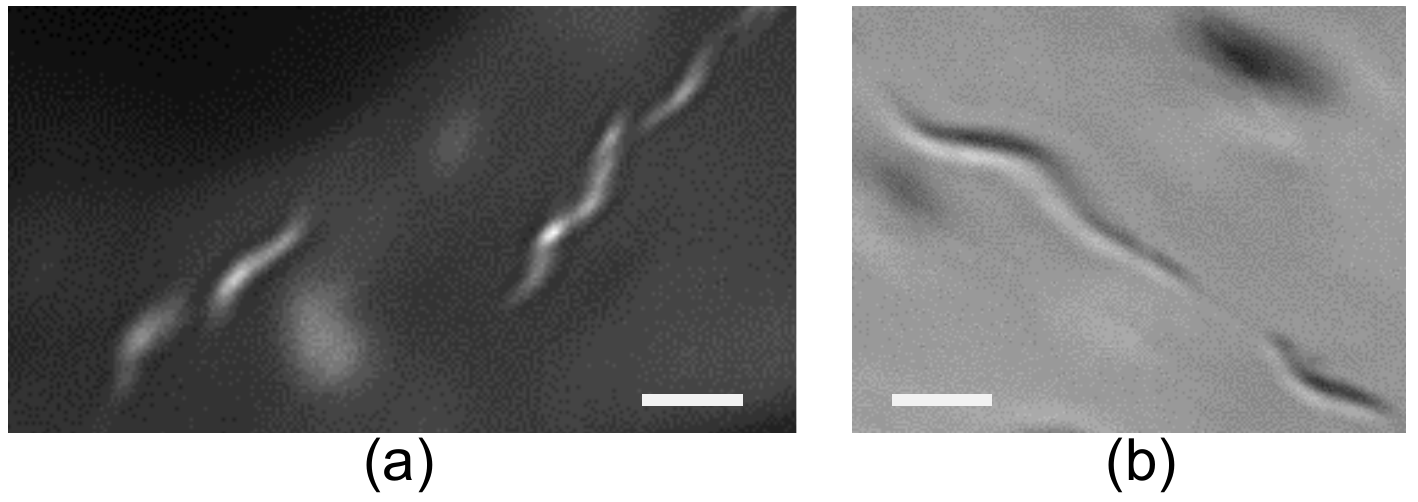}
	\caption{\textbf{Helical supramolecular structures} formed by condensation of filamentous viruses initially organized in a cholesteric mesophase, induced by depletion 
    interaction using poly(ethylene glycol) polymer (molecular weight $M_w$=2000~g/mol; Sigma-Aldrich) and observed by (a) polarizing and (b) differential interference contrast (DIC) microscopy. 
    Scale bar: 2~$\mu$m.
    }
	\label{HelicalBundles}
\end{figure}

\end{document}